\documentclass[11pt]{amsart}

\usepackage{amssymb}
\usepackage{amsmath}
\usepackage{epsfig}
\usepackage{comment}
\usepackage{subcaption}
\usepackage[section]{placeins}
\usepackage{mathrsfs}
\usepackage{graphicx}
\usepackage[notrig]{physics}
\usepackage{units}
\usepackage{color}

\usepackage{algorithm}
\usepackage{algorithmic}

\newtheorem{theorem}{Theorem}[section]

\makeatletter
\newcommand{\mbs}[1]{\boldsymbol{#1}}

\DeclareMathOperator{\mueq}{\overset{\mu}{\sim}}

\makeatletter \@addtoreset{figure}{section}
\def\thefigure{\thesection.\@arabic\c@figure}
\def\fps@figure{h, t}
\@addtoreset{equation}{section}

\makeatother

\newcommand{\pd}[2]{\frac{\partial #1}{\partial #2}}

\graphicspath{{./}{./Figures/}}

\begin{document}

\title[The energy-stepping Monte Carlo method]{The energy-stepping Monte Carlo method:\\
a Hamiltonian Monte Carlo method with a \\
100\% acceptance ratio}

\author{I.~Romero${}^{1,2}$ and M.~Ortiz${}^{3,4}$}

\address
{
    ${}^1$Universidad Polit\'ecnica de Madrid, Spain \\
    ${}^2$IMDEA Materials Institute, Spain \\
    ${}^3$California Institute of Technology, USA \\
    ${}^4$Hausdorff Center for Mathematics, Universit\"at Bonn, Germany
}

\begin{abstract}
We introduce the energy-stepping Monte Carlo (ESMC) method, a Markov chain Monte Carlo (MCMC) algorithm based on the conventional dynamical interpretation of the proposal stage but employing an energy-stepping integrator. The energy-stepping integrator is quasi-explicit, symplectic, energy-conserving, and symmetry-preserving. As a result of the exact energy conservation of energy-stepping integrators, ESMC has a 100\%\ acceptance ratio of the proposal states. Numerical tests provide empirical evidence that ESMC affords a number of additional benefits: the Markov chains it generates have weak autocorrelation, it has the ability to explore distant characteristic sets of the sampled probability distribution and it yields smaller errors than chains sampled with Hamiltonian Monte Carlo (HMC) and similar step sizes. Finally, ESMC benefits from the exact symmetry conservation properties of the energy-stepping integrator when sampling from potentials with built-in symmetries, whether explicitly known or not.
\end{abstract}

\maketitle


\section{Introduction}
\label{sec-intro}
Markov chain Monte Carlo (MCMC) methods are among the most important algorithms in scientific computing \cite{dongarra2000gr, diaconis2009yg, higham2016yg}. Introduced during World War II in the context of the Manhattan Project \cite{robert2011lu}, they have become indispensable in statistics, applied mathematics, statistical mechanics, chemistry, machine learning, and other fields (see, e.g., \cite{yeomans1992wf, sokal1996ti, liu2001mc, andrieu2003hi, diaconis2009yg}).

The goal of MCMC methods is to sample from an arbitrary probability distribution or, more generally, calculate expectations on (possibly unbounded) probability spaces with known densities. There exist well-known closed-form expressions that can perform this task when the distribution is simple or, more generally, when the sample space is one-dimensional \cite{madras2002uh}. However, this is not the case for complex and multidimensional sample distributions that are often of interest in practical applications. Many methods have been developed to alleviate computational costs and efforts still continue (see, for example, the monographs \cite{madras2002uh, robert2004gp, barbu2020is}).

\emph{Importance sampling} and \emph{rejection sampling} are among the simplest sampling methods for general probability distributions \cite{mackay2003tb}. It is possible to show that, under mild conditions, arbitrarily large samples can be obtained that are distributed according to a given probability distribution. However, in actual practice calculations can be exceedingly costly and inefficient and the applicability of those methods is limited to small and simple distributions.

Instead, MCMC methods are the \emph{de facto} choice in practical problems that require random sampling. The theory of these algorithms is grounded on the properties of Markov chains \cite{robert2004gp} and always involves a two-step process: given an existing sample, a new one is \emph{proposed} and it is either accepted or rejected before proceeding in a recursive fashion. The proposal step is key, and the success of an MCMC method mainly depends on its ability to efficiently generate samples that are accepted almost always, while simultaneously covering the sample space. Many variants of MCMC have been proposed in the pursuit of these goals, starting from the original Metropolis method \cite{metropolis1953mc} and including the popular modification by Hastings \cite{hastings1970fs}.

MCMC methods explore the sample space of a distribution and produce a sequence of samples that should be concentrated where the probability density is highest. This \emph{characteristic set} is often small and may consist of disconnected subsets that are far apart, with vast regions of low probability density separating them. The goal of MCMC methods is to explore as quickly as possible the characteristic set, covering all the regions where the probability measure is non-negligible. Indeed, when an MCMC method generates samples on regions with small probability most proposals are rejected, which adds to cost and renders the method inefficient. Thus, one of the goals of an MCMC method is to transition rapidly between high-probability regions of the characteristic set.

One particular class of MCMC algorithms that has proven particularly efficient is the class of \emph{Hamiltonian Monte Carlo} (HMC) methods. Originally introduced in the context of molecular dynamics, and initially referred to as the \emph{hybrid MC method} \cite{duane1987pw}, it exploits an interpretation of the sampling process as the motion of a generalized particle, of the type customarily considered in Hamiltonian mechanics \cite{neal2011mc, betancourt2018sg}. Such identification opens the door to the use of techniques developed to integrate in time Hamiltonian systems. Specifically, the proposal state in HMC employs \emph{symplectic} integrators, time-stepping algorithms designed to preserve some of the geometric structure of Hamiltonian systems. This strategy generates fast proposals that efficiently cover the characteristic set, as desired.

Symplectic integrators preserve the symplectic form of Hamiltonian systems and possibly other important invariants and symmetries. Their excellent properties have made them popular and have been exhaustively analyzed \cite{sanzserna1994wx}. One well-known result is that, even in conservative problems, symplectic integrators cannot preserve the total energy \cite{ge1988ul} unless the time step size is added to the unknowns of the problem \cite{kane1999vb}. This fact is exploited in MCMC methods: when the symplectic integration yields a proposal with a large energy error, this sample is rejected. A delicate balance is then sought: if the time integration of the MCMC method is performed for a small period of time or with a small time step, the proposal will most likely be accepted, albeit at great computational cost and time to traverse the characteristic set; conversely, if the time integration is performed for a long time or with a large time step, the method can potentially explore larger regions of the state space, albeit at the risk of proposing a sample that is likely to be rejected. Thus, the challenge HMC is to use fast, structure-preserving integrators capable of generating samples that efficiently explore the characteristic set while incurring in small energy errors that would entail their rejection.

The standard integrator for HMC is the leapfrog method \cite{leimkuhler2004uj}. Leapfrog is a first-order accurate, explicit, symplectic time integration scheme widely employed, e.~g., in molecular dynamics \cite{tuckerman2010wy}. Being explicit, the method is conditionally stable but has very low computational cost. In addition, as a result of its symplecticity it exactly preserves a \emph{shadow} Hamiltonian \cite{engle2005jc}, which limits the extent of energy drift from the exact value.

In this work, we propose a new class of MCMC methods of the HMC type, which we refer to as energy-stepping Monte Carlo (ESMC) methods, where the symplectic integrator (e.~g., leapfrog) is replaced by an energy-stepping integrator \cite{gonzalez2010hb, gonzalez2010uo}. These integrators are designed for Hamiltonian problems and possess remarkable properties such as symplecticity, unconditional stability, exact conservation of energy, and preservation of \emph{all} the symmetries of the Lagrangian, whether explicitly known or not. Energy-stepping integrators are essentially explicit, requiring the solution of \emph{one} scalar equation per integration step, irrespective of the dimension of the system. Given the remarkable properties of these integrators and their competitive cost, they suggest themselves as excellent candidates to replace other symplectic integrators in HMC: they can be expected to explore the characteristic set as efficiently as other symplectic integrators but, remarkably, propose samples with \emph{absolutely no rejections}. For an almost negligible increment in computational cost, because of its remarkable property of a 100\% acceptance rate ESMC has the potential to significantly improve the performance of HMC.

The article is structured as follows. In Section~\ref{sec-mcmc}, we review the concepts of MCMC methods, and provide the framework for ESMC. The Hamiltonian Monte Carlo method is presented in Section~\ref{sec-hmc}, and the role played by the time integration step is carefully stressed. Next, Section~\ref{sec-es} describes the energy-stepping time integration scheme, without any reference to Monte Carlo methods but, rather, as a method designed to integrate Hamiltonian mechanics. Then, in Section~\ref{sec-esmc}, we introduce the main contribution of this work, namely, the ESMC method. Numerical simulations that illustrate its performance and the comparison with other standard MCMC methods are presented in Section~\ref{sec-examples}. We emphasize that the numerical tests presented in this work focus on exemplifying the fundamental properties of ESMC and are not intended to be representative of production-ready codes or libraries such as STAN and NIMBLE (e.~g.~\cite{hoffman2011ns}), which are the result of extensive development and fine tuning. Finally, Section~\ref{sec-summary} closes the article with a summary of the main findings and outlook for further work.

\section{Markov chain Monte Carlo methods}
\label{sec-mcmc}
By way of background, we start by briefly reviewing the fundamentals of Markov chains and their link with sampling methods. In many applications of statistics, it is necessary to evaluate expectations relative to probability density functions that are complex and, therefore, impossible to obtain analytically. These situations appear, for instance, when doing inference in Bayesian models or, simply, when predicting the output of random processes. In view of the difficulty in obtaining closed-form expectations, numerical approximation is required.

The standard procedure is as follows. Let $\Omega$ be an $n-$dimensional sample space, not necessarily compact, and $\mu$ a Lebesgue-continuous probability measure
\begin{equation}
  \label{eq-prob-measure}
  \mu(B) = \int_B \pi(x) \,\mathrm{d} \Omega \,,
\end{equation}
for all Lebesgue-measurable sets $B\subset\Omega$ and integrable probability density function $\pi:\Omega\to \mathbb{R}^+$. The main problem is to calculate
\begin{equation}
  \label{eq-expectation}
  \mathbb{E}_{\mu}[f] = \int_{\Omega} f(x)\; \pi(x) \,\mathrm{d} \Omega\,,
\end{equation}
for all bounded continuous functions $f$. In order to approximate integral~\eqref{eq-expectation}, assume that we have a collection $\{x_k\}_{k=1}^N$ of independent, identically distributed samples with probability $p$. Then, the corresponding empirical approximation of the expectation is
\begin{equation}
  \label{eq-expectation-approx}
  \mathbb{E}_{\mu}[ f]
  \approx
  S_N
  :=
  \frac{1}{N}
  \sum_{k=1}^{N} f(x_k)\,.
\end{equation}
By the weak-* density of Diracs in the space of probability measures, possibly under moment constraints (e.~g., \cite{ambrosio:2008, Villani:2009}), it is possible to chose (unbiased) sequences $\{x_k\}_{k=1}^N$ such that
%
\begin{equation}
  \label{eq-statistic}
  \lim_{N\to\infty} S_N = \mathbb{E}_{\mu}[f]\,,
\end{equation}
for every bounded continuous function $f$.

Evidently, the key to calculating converging empirical expectations of the form (\ref{eq-expectation-approx}) efficiently is the construction of the sample array $\{x_k\}_{k=1}^N$. Naive methods to generate samples using, for example, rejection sampling are extremely costly, especially in high-dimension sample spaces \cite{mackay2003tb}.

The standard workhorse for this task is the Markov chain Monte Carlo method (MCMC) and its variants \cite{mackay2003tb}. MCMC methods are designed to generate sequences of samples distributed according to the target probability and spending most of the computational effort in those regions of high probability density. To this end, special random walks in sample space are generated with stochastic rules determining whether some of these states should be discarded or not.

\subsection{Markov chains}
\label{subs-markov}
We summarize the basic concepts of Markov chains that are required to define MCMC methods and we refer to specialized monographs for additional results and proofs (e.g., \cite{robert2004gp}).

A (discrete) Markov chain is a finite sequence $\{q_k\}_{k=1}^N$ of random variables that possesses the \emph{Markov property}: the conditioned probability of $q_{k+1}$ given all past values $q_k,q_{k-1},\ldots q_1,q_0$ is actually a function of $q_k$ only. This probability is known as the \emph{transition kernel}~$K$ and we write
\begin{equation}
  \label{eq-markov}
  q_{k+1} \vert q_k, q_{k-1}, \ldots, q_1,q_0
  \sim
  q_{k+1} \vert q_k
  \sim
  K(q_{k}, q_{k+1})\,.
\end{equation}

The Markov chains of interest for MCMC methods are those that are \emph{irreducible} and possess a \emph{stationary distribution}. The first property, irreducibility, requires that, given any initial value $q_0$ and an arbitrary non-empty set, the probability that --- at some time -- the Markov chain will generate a state belonging to the set is greater than zero. The second property, the existence of a stationary distribution, demands that there exist a probability density function $\pi$ that is preserved by the transition kernel, i.~e., if $q_k\sim \pi$ then $q_{k+1}\sim \pi$, or, equivalently,
\begin{equation}
  \label{eq-stationary}
  \int K(x,y)\, \pi(x) \; \mathrm{d} x = \pi(y)\,.
\end{equation}
A necessary condition for a Markov chain to be stationary is that it is irreducible.

In a recurrent Markov chain, the stationary distribution $\pi$ is limiting, i.~e., for almost any initial sample $q_0$ the sample $q_k$ is distributed as $\pi$ for large enough $k$. This property --- also referred to as \emph{ergodicity} --- is exploited in the formulation of MCMC methods: the search of samples distributed according to~$\pi$ aims at building an ergodic Markov chain with a transition kernel~$K$ and stationary probability identical to~$\pi$. If these properties are attained, the values of the chain will eventually be distributed as unbiased samples from~$\pi$.

A stationary Markov chain is \emph{reversible} if $K(x,y) = K(y,x)$. Moreover, a Markov chain satisfies the \emph{detailed balance condition} if there exists a probability $\pi$ such that
\begin{equation}
  \label{eq-detailed-balance}
  K(y,x)\, \pi(y) = K(x,y)\, \pi(x)
\end{equation}
If the transition kernel of a Markov chain verifies the detailed balance condition with function $\pi$, then the latter is a stationary probability associated with the transition kernel $K$ and the chain is reversible. While this condition is sufficient but not necessary for the two properties to hold,
detailed balance is easy to verify and is often employed in practice to analyze MCMC methods.

\subsection{Markov chains for Monte Carlo methods}
The key idea behind MCMC methods is that independent, identically distributed unbiased samples with probability distribution $\pi$ can be effectively obtained simply by collecting the states in a ergodic Markov chain defined by a transition kernel~$K$ \emph{designed} to verify the detailed balance condition~\eqref{eq-detailed-balance} with probability~$\pi$. However, this strategy leaves considerable freedom of sampling from $\pi$ in many different ways, depending on the Markov chain employed.

Metropolis-Hastings MCMC, the most common type of MCMC method, employs a \emph{proposal} or \emph{instrumental} probability distribution $p$ to define the transition kernel. Specifically, and given a chain with current value $q_k$, the next state is obtained by a two-step procedure: first, a random sample $\tilde{q}$ is generated with probability density $p(\tilde{q}\vert q_k)$. Then, the state $q_{k+1}$ is selected to be equal to $\tilde{q}$ with probability $\rho(q_k,\tilde{q})$, or equal to the previous state $q_k$ with probability $1-\rho(q_k,\tilde{q})$, where
\begin{equation}
  \label{eq-rho}
  \rho(q_k,\tilde{q})
  =
  \min \left\{
  \frac{\pi(\tilde{q})}{\pi(q_{k})} \frac{p(q_{k}\vert \tilde{q})}{p(\tilde{q}\vert q_k)} , 1
  \right\}\,.
\end{equation}
This transition map can be shown to satisfy the detailed balance condition for all probabilities $p$ and $\pi$ (cf.~\cite{robert2004gp}). It bears emphasis that in the Metropolis-Hastings algorithm, once a proposal $\tilde{q}$ is generated, it may be rejected. Thus, the ratio of the accepted to the rejected proposal samples can dramatically affect the ability of the method to explore the characteristic set of the target probability distribution~$\pi$, as well as its computational cost \cite{betancourt2018sg}. This is a delicate tradeoff for which there exist heuristic rules: too small an acceptance ratio leads to an inefficient sampling strategy, whereas a too-large one might indicate that the Markov chain is covering the characteristic set of $\pi$ too slowly.

The simplest variant of MCMC, commonly known as \emph{random walk} MCMC, employs a proposal distribution that is a Gaussian centered at the current value of the Markov chain. This proposal is obviously symmetric, and its variance can be selected to optimize the acceptance ratio. Other proposals, such as the Gamma or the Student-t distributions, may be exploited to bias the new states of the Markov chain away from previously explored regions.

\section{Hamiltonian Monte Carlo methods}
\label{sec-hmc}
Hamiltonian Monte Carlo methods (HMC) are yet another family of algorithms designed to sample complex target probability distributions. They have proven advantageous over other, more traditional, MCMC methods, especially for high-dimensional sample spaces such as appear in problems of statistical mechanics. Originally introduced in the context of molecular dynamics and dubbed the \emph{hybrid Monte Carlo method} \cite{duane1987pw}, this family of algorithms exploits ideas and numerical methods used in the study of dynamical systems, and more specifically, Hamiltonian problems on symplectic or Poisson spaces~\cite{neal2011mc}.

Similarly to Metropolis-Hastings MCMC, HMC methods generate Markov chains of independent samples distributed according to a target probability distribution, but differ thereof in that they take advantage of the geometrical information provided by the \emph{gradient} of the target probability density. By interpreting the state of the Markov chain as a particle moving conservatively under the action of a fictitious potential defined by the target probability, HMC methods generate chains that have been shown to efficiently explore the regions of high probability density, even when highly concentrated~\cite{betancourt2018sg}. Thus, HMC methods essentially replace the \emph{proposal} step of any MCMC method but keep the accept/reject step unchanged, albeit in a form more conveniently expressed in terms of the surrogate energy.

To describe the HMC approach, let $q$ denote as before the random variable for which a target probability with density $\pi$ is known. The goal of HMC is to construct a collection $\{q_k\}_{k=1}^N$ of independent samples identically distributed according to $\pi$. Let us assume that
\begin{equation}
  \label{eq-hmc-p}
  \pi(q) = \frac{1}{Z_q} \mathrm{e}^{-V(q)}\,,
\end{equation}
where we refer to $V:\Omega\to \mathbb{R}$ as the \emph{potential function}
and $Z_q$ is a normalizing factor. We additionally introduce an ancillary random variable $p:\Omega\to \mathbb{R}$ and postulate that the two
random variables $q$ and $p$, defining coordinate $z=(q,p)$ in \emph{phase space}, satisfy
\begin{equation}
  \label{eq-joint}
  (q,p) \sim \pi(q,p) = \pi(p \vert q)\, \pi(q)\,,
\end{equation}
with
\begin{equation}
  \label{eq-p-distribution}
  \pi(p \vert q) = \frac{1}{Z_p} \textrm{e}^{-K(q,p)}
\end{equation}
and $K:\Omega\times\Omega\to \mbs{R}^+$ referred to as the \emph{kinetic energy}. Then, the joint distribution~\eqref{eq-joint} follows as
\begin{equation}
  \label{eq-joint2}
  \pi(q,p) = \frac{1}{Z_{q}\;Z_p} \textrm{e}^{-V(q)-K(q,p)} = \frac{1}{Z_{H}}\textrm{e}^{-H(q,p)}\,,
\end{equation}
where the normalizing constant is now $Z_{H}=Z_p\,Z_q$ and
\begin{equation}
\label{eq-hamiltonian}
H(q,p) = V(q) + K(q,p)
\end{equation}
is the \emph{Hamiltonian}. Note that the marginal distribution of the pair $(q,p)$ with respect to $p$ satisfies
\begin{equation}
  \label{eq-margian}
  \int_{\mathbb{R}^n} \pi(q,p) \,\mathrm{d} p =
  \int_{\mathbb{R}^n}  \pi(p\vert q) \, \pi(q) \,\mathrm{d} p = \pi(q)\,.
\end{equation}
Hence, if a sequence $\{(q_k,p_k)\}_{k=1}^N$ is sampled with distribution $\pi(q,p)$, the submersion $(q_k,p_k)\mapsto q_{k}$ is distributed with probability $\pi(q)$.

\subsection{Algorithmic details}
\label{subs-hmc-details}

Similarly to other MCMC methods, HMC proceeds recursively, in the process generating a Markov chain of states. On each iteration, a state is proposed and then either accepted or rejected. Specifically, let $q_k$ be the last accepted state of the chain. Then, to compute the next state, HMC methods calculate the following:
\begin{enumerate}
\item In the first step, a random value for the momentum $p_k$ is sampled with probability distribution $\pi(p_k\vert q_k)$ as defined in Eq.~\eqref{eq-p-distribution}. To that end, a kinetic potential is selected. A simple choice is
  \begin{equation}
    \label{eq-kine1}
    K(q,p) = \frac{1}{2} p\cdot M^{-1} p + \log |M|\,,
  \end{equation}
with $M$ a constant metric. Other more sophisticated kinetic energies make use of metrics $M$ that depend on the configuration in an attempt to better capture geometrical detail, but such extensions are not considered here.

\item In the second step, the pair $z_k=(q_k,p_k)$ evolves under a Hamiltonian flow
\begin{equation}
  \label{eq-hamiltons}
  \dot{q}(t) = \pd{H}{p}(q(t),p(t))\,,
  \qquad
  \dot{p}(t) = - \pd{H}{q}(q(t),p(t))\,,
\end{equation}
for a time interval $t\in[0,T]$, and with initial conditions
\begin{equation}
  q(0) = q_k\,,
  \qquad
  p(0) = p_k\,.
\end{equation}
Since the potential may be an arbitrary function, an exact solution to this evolution problem is not available in general and a time integrator, such as leapfrog, needs to be used instead. A delicate tradeoff concerns the choice of the length $T$ of the integration interval and its relationship with the time step size required for stability.

\item The proposal state $\tilde{z}=(\tilde{q},\tilde{p})=(q(T),p(T))$ is accepted with probability
\begin{equation}
    \rho(z_k,\tilde{z}) = \min
    \left\{
    1, \exp[H(z_k)-H(\tilde{z})]
    \right\} ,
\end{equation}
and rejected with probability $1-\rho(z_k, \tilde{z})$ otherwise. If accepted, we set $z_{k+1}=(q_{k+1},p_{k+1}) = \tilde{z}$ and the state $q_{k+1}$ is added to the Markov chain.
\end{enumerate}

\section{Energy-stepping integrators}
\label{sec-es}

We proceed to review \emph{energy-stepping} integrators, a paradigm that differs fundamentally from classical integrators such as Runge-Kutta, multistep, or one-leg methods \cite{hairer1987vh, hairer1996ub}, but shares all the advantageous features of symplectic methods as well as possessing some unique ones \cite{gonzalez2010hb, gonzalez2010uo}.

We begin by focusing on systems characterized by Lagrangians $L: \mathbb{R}^N \times \mathbb{R}^N \to \mathbb{R}$ of the form
\begin{equation}\label{eq:ES:L}
    L(q,\dot{q})
    =
    \frac{1}{2} \dot{q}^T M \dot{q} - V(q) \,,
\end{equation}
where $M$ is a mass matrix and $V(q)$ is the potential energy function. The classical motivation for systems of this type comes from celestial, structural, solid, and molecular mechanics \cite{arnold1989wt}. This framework fits within HMC since the Lagrangian (\ref{eq:ES:L}) is equivalent to the Hamiltonian~\eqref{eq-hamiltonian} through the
\begin{equation}
    \label{eq-legendre}
    L(q,\dot{q}) = \sup_p \left( \dot{q}\,p - H(q,p) \right)\,.
\end{equation}
The trajectories of the Lagrangian~\eqref{eq:ES:L} render stationary Hamilton's action
\begin{equation}
\label{eq-action}
    I = \int L(q(t),\dot{q}(t))\; dt\,.
\end{equation}
Since such trajectories are not easy to find, the classical approximation paradigm is to discrete the action in time, leading to variational time integrators.

The energy-stepping paradigm is at variance with time-stepping in that it employs {\sl exact solutions} of an {\sl approximate Lagrangian} that can be solved exactly. For Lagrangians of the form (\ref{eq:ES:L}), \cite{gonzalez2010hb, gonzalez2010uo} propose the approximate Lagrangians
\begin{equation}\label{eq:ES:Lh}
    L_h(q,\dot{q})
    =
    \frac{1}{2} \dot{q}^T M \dot{q} - V_h(q)\,,
\end{equation}
where $V_h$ is some exactly solvable approximation of the potential energy. The energy-stepping method specifically considers {\sl piecewise constant} approximations of the potential energy, i.~e., \emph{terraced} approximations $V_h$ of constant energy-step height $h$ defined as
\begin{equation}\label{eq:Vh2}
    V_h(q)
    =
    h \lfloor h^{-1} V(q) \rfloor\,,
\end{equation}
where $\lfloor\cdot\rfloor$ is the floor function, i.~e., $\lfloor x \rfloor = \max \{ n \in \mathbb{Z} : n \leq x\}$. Based on this definition, $V_h$ is the largest piecewise-constant function with values in $h\mathbb{Z}$ majorized by $V$.

\begin{figure}[t]
\centering{
    \includegraphics[width=0.49\textwidth]{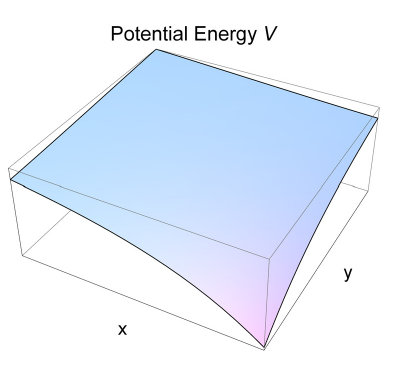}
    \includegraphics[width=0.49\textwidth]{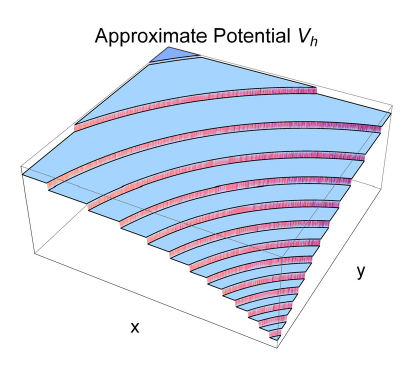}
    \includegraphics[width=0.49\textwidth]{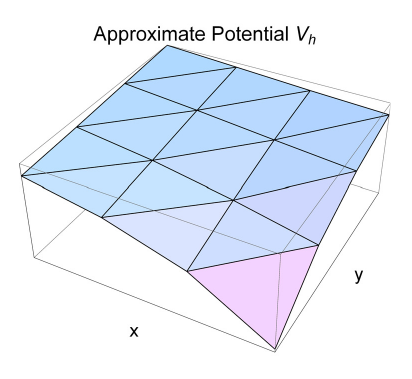}
}
\caption{Kepler problem. Exact, piecewise constant and piecewise linear continuous approximate potential energies. }\label{Fig-KeplerPotential}
\end{figure}

An approximating sequence of potential energies, and by extension Lagrangians, can be built by selecting energy steps of decreasing height. Other types of approximations, such as piecewise linear interpolations of the potential energy, also result in exactly integrable approximating systems \cite{gonzalez2010hb} but will not be considered here for definiteness. See Fig.~\ref{Fig-KeplerPotential} for an illustration of a piecewise constant and a piecewise linear approximation of the Kepler potential.

\subsection{Computation of the exact trajectories of the approximating Lagrangian} \label{subsec:exacttrajectories}

Following \cite{gonzalez2010uo}, we describe next the calculation of the \emph{exact} trajectories generated by the terraced Lagrangians $L_{h}\left(q,\dot{q}\right)$.

\begin{figure}[t]
\centering{
    \includegraphics[width=\textwidth]
    {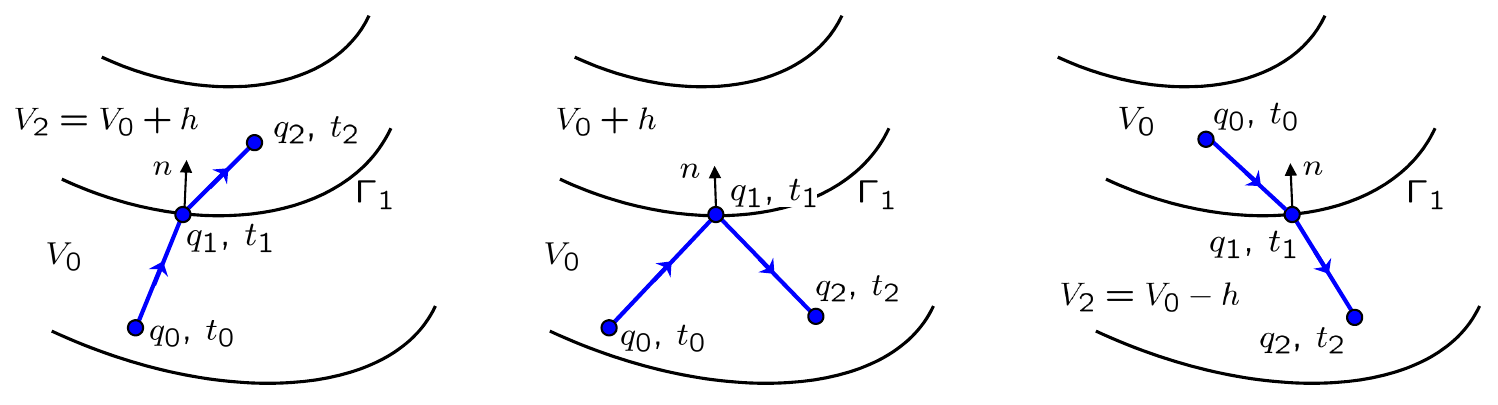}
}
\caption{Trajectories of a system with a piecewise constant potential energy. Left: uphill diffraction step; Center: uphill reflection step; Right: downhill diffraction step.}
\label{Fig-PossibleTrajectories}
\end{figure}

Suppose that a mechanical system is in configuration $q_{0}$ at time $t_{0}$, in configuration $q_{2}$ at time $t_{2}$, and that during the time interval $[t_{0},t_{2}]$ it intersects one single jump surface $\Gamma_{1}$ separating two regions of constant energies $V_0$ and $V_2$ (see Figure~\ref{Fig-PossibleTrajectories}). Based on the form of $V_h$, $\Gamma_1$ is the level surface $V = V_2$ for an uphill step $V_2 = V_0+h$, or the level surface $V = V_0$ for a downhill step, $V_2 = V_0-h$. For simplicity, we shall further assume that $V$ is differentiable and that all energy-level crossings are transversal, i.~e.,
\begin{equation}\label{eq:transversal}
    n(q_1)\cdot\dot{q}_1^- \neq 0
\end{equation}
where $\dot{q}_{1}^{-}=\dot{q}\left(t_{1}^{-}\right)$ and $n(q_1)$ is a vector normal to $\Gamma_{1}$ pointing in the direction of advance.

Under these assumptions, the action integral~\eqref{eq-action} over the time interval $[t_{0},t_{2}]$ follows as
\begin{equation}\label{eq:discreteaction}
    I_{h}
    =
    \int\nolimits_{t_{0}}^{t_{2}}L_{h}\left(q,\dot{q}\right)~dt
    =
    \int\nolimits_{t_{0}}^{t_{1}}L_{h}(q,\dot{q})~dt +
    \int\nolimits_{t_{1}}^{t_{2}}L_{h}(q,\dot{q})~dt
\end{equation}
where $t_{1}$ is the time at which the trajectory intersects $\Gamma_{1}$. In regions where $V_{h}(q)$ is constant the trajectory $q(t)$ is linear in time. Therefore, the action of the system can be computed \emph{exactly} and reduces to
\begin{equation}
\begin{split}
    I_{h}
    & =
    \left(t_{1}-t_{0}\right)  \left\{
    \frac{1}{2}\left(\frac{q_{1}-q_{0}}{t_{1}-t_{0}}\right)^{T}M\left( \frac{q_{1}-q_{0}}{t_{1}-t_{0}}\right)  -V_0
    \right\}
    \\ & +
    \left(t_{2}-t_{1}\right)  \left\{
    \frac{1}{2}\left(\frac{q_{2}-q_{1}}{t_{2}-t_{1}}\right)^{T}M\left( \frac{q_{2}-q_{1}}{t_{2}-t_{1}}\right)  -V_2
    \right\}\,,
\end{split}
\end{equation}
where $q_{1}=q\left(t_{1}\right)$ is constrained to be on the jump surface $\Gamma_{1}$. Assuming differentiability of $\Gamma_{1}$, stationarity of the action $I_h$ with respect to $(t_{1},q_{1})$ additionally gives the energy conservation equation
\begin{equation}
    \left(\frac{q_{1}-q_{0}}{t_{1}-t_{0}}\right)^{T}M\left(\frac{q_{1}
    -q_{0}}{t_{1}-t_{0}}\right)  +2\,V_0
    =
    \left(\frac{q_{2}-q_{1}}{t_{2}-t_{1}}\right)^{T}M\left(\frac{q_{2}
    -q_{1}}{t_{2}-t_{1}}\right) + 2\,V_2\,,
\label{Eqn-StationaryAction1}
\end{equation}
and the linear momentum balance equation
\begin{equation}
    M\frac{q_{1}-q_{0}}{t_{1}-t_{0}}-
    M\frac{q_{2}-q_{1}}{t_{2}-t_{1}}+
    \lambda\, n(q_{1})
    = 0\,,
\label{Eqn-StationaryAction2}
\end{equation}
where $\lambda$ is a Lagrange multiplier.

In order to make contact with time-integration schemes, we reformulate the problem slightly by assuming that $t_{0}$, $q_{0}$ --- the latter on a jump surface $\Gamma_{0}$ except, possibly, at the initial time --- and the initial velocity
\begin{equation}
    \dot{q}_{0}^{+}
    =
    \dot{q}\left(t_{0}^{+}\right)
    =
    \frac{q_{1}-q_{0}}{t_{1}-t_{0}}
\end{equation}
are known. Let $t_{1}$ and $q_{1}$ be the time and point at which the trajectory intersects the next jump surface $\Gamma_{1}$. We then seek to determine
\begin{equation}
    \dot{q}_{1}^{+}
    =
    \dot{q}\left(t_{1}^{+}\right)\,.
\end{equation}
A reformulation of Eqs.~\eqref{Eqn-StationaryAction1} and \eqref{Eqn-StationaryAction2} in terms of $\dot{q}_{1}^{+}$ gives
\begin{align}
    \left(\dot{q}_{1}^{+}\right)^{T}M\dot{q}_{1}^{+}
    & =
    \left(\dot{q}_{1}^{-}\right)^{T}M\dot{q}_{1}^{-}-2\Delta V
    \label{Eqn-StationaryAction1b}
    \\
    \dot{q}_{1}^{+}
    & =
    \dot{q}_{1}^{-}+\lambda M^{-1}n(q_{1})\,,
    \label{Eqn-StationaryAction2b}
\end{align}
where $\dot{q}_{1}^{-} = \dot{q}_{0}^{+}$ and the potential energy jump is $\Delta V = V_{h}\left(q(t_1^+)\right) - V_{h}\left(q(t_1^-)\right)$. Next, we proceed to examine the various alternatives that can arise in the solution of \eqref{Eqn-StationaryAction1b} and \eqref{Eqn-StationaryAction2b}.

\subsubsection{Diffraction by downhill energy step}\label{subsubsec:downhill}

Suppose that $\Delta V = -h$, i.~e., the system decreases its potential energy as the trajectory crosses $\Gamma_{1}$. Then (\ref{Eqn-StationaryAction1b}) becomes
\begin{equation}\label{Eqn-StationaryAction1c}
    \left(\dot{q}_{1}^{+}\right)^{T}M\dot{q}_{1}^{+}
    =
    \left(\dot{q}_{1}^{-}\right)^{T}M\dot{q}_{1}^{-}
    +
    2 h\,.
\end{equation}
In this case, the system of equations \eqref{Eqn-StationaryAction2b} - \eqref{Eqn-StationaryAction1c} has a real solution
\begin{equation}\label{eq:qplusi}
    \dot{q}_{1}^{+}
    =
    \dot{q}_{1}^{-}
    +
    \frac{-\dot{q}_{1}^{-}\cdot n_{1}
    -
    \sqrt
    {\left(\dot{q}_{1}^{-}\cdot n_{1}\right)^{2}
    +
    2 h \, n_{1}^{T}M^{-1}n_{1}}}
    {n_{1}^{T}M^{-1}n_{1}}
    M^{-1}n_{1}\,,
\end{equation}
with $n_1=n(q_1)$ This solution represents the diffraction, or change of direction, of the trajectory by a downhill energy step.

\subsubsection{Diffraction by uphill energy step}\label{subsubsec:uphill}

Suppose now that $\Delta V = h$, i.~e., the system increases its potential energy as the trajectory crosses $\Gamma_{1}$. Then, (\ref{Eqn-StationaryAction1b}) becomes
\begin{equation}\label{Eqn-StationaryAction1d}
    \left(\dot{q}_{1}^{+}\right)^{T}M\dot{q}_{1}^{+}
    =
    \left(\dot{q}_{1}^{-}\right)^{T}M\dot{q}_{1}^{-}
    -
    2 h \,.
\end{equation}
Additionally, suppose that
\begin{equation}\label{eq:qplustrans}
    \left(\dot{q}_{1}^{-}\cdot n_{1}\right)^{2}
    >
    2 h \, n_{1}^{T}M^{-1}n_{1}\,.
\end{equation}
Then, Eqs. \eqref{Eqn-StationaryAction2b} and \eqref{Eqn-StationaryAction1d} again has a real solution, namely,
\begin{equation}\label{eq:qplusii}
    \dot{q}_{1}^{+}
    =
    \dot{q}_{1}^{-}
    +
    \frac{-\dot{q}_{1}^{-}\cdot n_{1}
    +
    \sqrt
    {\left(\dot{q}_{1}^{-}\cdot n_{1}\right)^{2}
    -
    2 h \, n_{1}^{T}M^{-1}n_{1}}}
    {n_{1}^{T}M^{-1}n_{1}}
    M^{-1}n_{1}\,.
\end{equation}
This solution represents the diffraction of the trajectory by an uphill energy step when the system has sufficient initial energy to overcome the energy barrier.

\subsubsection{Reflection by uphill energy step}\label{subsubsec:bounce}

Suppose now that $\Delta V = h$, i.~e., the system increases its potential energy as the trajectory crosses $\Gamma_{1}$, but, contrary to the preceding case,
\begin{equation}
    \left(\dot{q}_{1}^{-}\cdot n_{1}\right)^{2}
    <
    2 h \, n_{1}^{T}M^{-1}n_{1}\,.
\end{equation}
Then, the system \eqref{Eqn-StationaryAction2b}-\eqref{Eqn-StationaryAction1d} has no real solutions, indicating that the mechanical system does not have sufficient energy to overcome the energy barrier. Instead, the trajectory remains within the same potential energy level and equation \eqref{Eqn-StationaryAction1b} becomes
\begin{equation}\label{Eqn-StationaryAction1e}
    \left(\dot{q}_{1}^{+}\right)^{T}M\dot{q}_{1}^{+}
    =
    \left(\dot{q}_{1}^{-}\right)^{T}M\dot{q}_{1}^{-}\,.
\end{equation}
In this situation, the system of equations \eqref{Eqn-StationaryAction2b} - \eqref{Eqn-StationaryAction1e} has the solution
\begin{equation}\label{eq:qplusiii}
    \dot{q}_{1}^{+}
    =
    \dot{q}_{1}^{-}
    -
    2\frac{\dot{q}_{1}^{-}\cdot n_{1}}{n_{1}^{T}M^{-1}n_{1}} M^{-1}n_{1}\,.
\end{equation}
This solution represents the reflection of the trajectory by an uphill potential energy step when the system does not have sufficient initial energy to overcome the energy barrier.

\subsection{Summary of the energy-stepping scheme}
\label{subsec:energystep}

We proceed to summarize the relations obtained in the foregoing and define the \emph{energy-stepping} integrator resulting from a piecewise-constant approximation of the potential energy.

Suppose that $\left(t_{k} ,q_{k},\dot{q}_{k}^{+}\right)$ and a piecewise-constant approximation of the potential energy $V_{h}$ are given. Let $t_{k+1}$ and $q_{k+1}$ be the time and point of exit of the rectilinear trajectory $q_{k} + (t-t_{k}) \dot{q}_{k}^{+}$ from the set $\{V = h\mathbb{Z}\}$. Let $\Delta V$ be the jump of the potential energy at $q_{k+1}$ in the direction of advance. The, the updated velocity is
\begin{equation}
    \dot{q}_{k+1}^{+}
    =
    \dot{q}_{k}^+ + \lambda_{k+1}M^{-1}n_{k+1}\,,
\end{equation}
where $n_{k+1}= n(q_{k+1})$ and
\begin{equation}
    \lambda_{k+1}
    =
    \left\{
\begin{array}[c]{ll}
    -2\frac{\dot{q}_{k}^{+}\cdot n_{k+1}}{n_{k+1}^{T}M^{-1}n_{k+1}},
    \qquad\text{if }\left(\dot{q}_{k}^{+}
    \cdot
    n_{k+1}\right)^{2} < 2\Delta V
    \left(n_{k+1}^{T}M^{-1}n_{k+1}\right),
    & \\
    \frac{-\dot{q}_{k}^{+}
    \cdot
    n_{k+1}
    +
    \operatorname{sign}
    \left(\Delta V\right)
    \sqrt{\left(\dot{q}_{k}^{+}
    \cdot
    n_{k+1}\right)^{2}
    -
    2\Delta V\left(n_{k+1}^{T}M^{-1}n_{k+1}\right)}}
    {n_{k+1}^{T}M^{-1}n_{k+1}},
    \quad \text{otherwise}.
\end{array}
\right.
\end{equation}
These relations define a discrete propagator
\begin{equation}
\label{eq-propagator}
    \Phi_{h}:
    \left(t_{k},q_{k},\dot{q}_{k}^{+}\right)
    \mapsto
    \left(t_{k+1},q_{k+1},\dot{q}_{k+1}^{+}\right)
\end{equation}
that can be applied recursively to generate a discrete trajectory.

\begin{algorithm}
\caption{Energy-stepping scheme}
\label{Alg-EnergyStepping}
\begin{algorithmic}[1]
\REQUIRE $V(q)$, $q_{0}$, $\dot{q}_{0}$, $t_{0}$, $t_{f}$ and the energy step $h $
\STATE $i \leftarrow 0$
\WHILE{$t_{i}<t_{f}$}
\STATE $t_{i+1}\leftarrow $ {\sc Smallest-Root}$\left( V\left(q_{i}+\left(t_{i+1}-t_{i}\right) \dot{q}_{i}\right) -V\left(q_{i}\right) +\Delta V = 0 \right)$
\STATE $q_{i+1}\leftarrow q_{i}+\left( t_{i+1}-t_{i}\right) \dot{q}_{i}$
\STATE $n_{i+1}\leftarrow \nabla V\left( q_{i+1}\right)$
\STATE $\dot{q}_{i+1}\leftarrow$ {\sc Update-Velocities}$\left( \dot{q}_{i},n_{i+1},h \right)$
\STATE $i \leftarrow i+1$
\ENDWHILE
\IF{$t_i>t_f$}
\STATE $q_i \leftarrow (1-\frac{t_f - t_{i-1}}{t_i-t_{i-1}}) q_{i-1} + \frac{t_f - t_{i-1}}{t_i-t_{i-1}} q_i$
\STATE $t_i \leftarrow t_f$
\ENDIF
\end{algorithmic}
\end{algorithm}

The computational workflow of the energy-stepping scheme is summarized in Algorithm \ref{Alg-EnergyStepping}. The algorithm combines two methods. The first method {\sc Smallest-Root} determines the root $t_{i+1}>t_i$ of the equation
\begin{equation}\label{eq:SmallestRoot}
    V\left(q_{i}+\left(t_{i+1}-t_{i}\right)  \dot{q}_{i}\right)-V\left(q_{i}\right)
    +
    \Delta V=0\,,
\end{equation}
where $\Delta V$ can take values in $\left\{ 0,h\right\}$. The second method {\sc Update-Velocities} updates velocities and reduces to only two situations. The method is defined in Algorithm \ref{Alg-UpdateVelocities}.

\begin{algorithm}
\caption{{\sc Update-Velocities}$\left ( \dot{q}_{0},n_{1},h \right )$}
\label{Alg-UpdateVelocities}
\begin{algorithmic}[1]
\IF{$\dot{q}_{0}\cdot n_{1}\geq 0$}
\STATE $\Delta V \leftarrow h$ \qquad
\ELSE
\STATE $\Delta V \leftarrow -h$ \qquad
\ENDIF
\IF{$\left( \dot{q}_{0}\cdot n_{1}\right) ^{2}\leq 2\Delta V \left( n_{1}^{T}M^{-1}n_{1}\right) $}
\STATE $\lambda \leftarrow -2\frac{\dot{q}_{0}\cdot n_{1}}{n_{1}^{T}M^{-1}n_{1}}$
\ELSE
\STATE $\lambda \leftarrow \frac{-\dot{q}_{0}\cdot n_{1} +
\operatorname{sign}\left( \Delta V \right) \sqrt{\left( \dot{q}_{0}\cdot n_{1}\right) ^{2}-2\Delta V \left( n_{1}^{T}M^{-1}n_{1}\right) }}{n_{1}^{T}M^{-1}n_{1}}$
\ENDIF
\STATE \bf{return} $\dot{q}_{0} + \lambda M^{-1}n_{1}$
\end{algorithmic}
\end{algorithm}

Remarkably, the energy-stepping method does not require the solution of a system of equations and, therefore, its complexity is comparable to that of explicit methods. However, the need to compute the root of a nonlinear scalar function per step adds somewhat to the overhead of the algorithm.

\subsection{Properties of energy-stepping integrators}
\label{Section-PreservingProperties}

We summarize from \cite{gonzalez2010hb, gonzalez2010uo} the main properties of energy-stepping integrators

The terraced approximation of the potential energy (\ref{eq:Vh2}) preserves all the symmetries of the system. By Noether's theorem, and since the discrete trajectories are exact trajectories of a Lagrangian system, the energy-stepping method conserves all the momentum maps and the symplectic structure of the original Lagrangian system. In particular, energy can be viewed as the momentum map associated with the time-reparametrization of a Lagrangian defined in space-time. The energy-stepping method must also preserve this symmetry, and thus, it is exactly energy-conserving for all step heights.

In summary, the energy-stepping integrator combines the following properties:
\begin{enumerate}
    \item It exactly preserves the symplectic form.
    \item It exactly preserves all the momentum maps that originate from symmetries of the Lagrangian, including the linear and angular momenta, and the energy.
    \item It is time-reversible.
    \item Its solution converges when $h\to0$ to the trajectory that makes stationary the exact action~\eqref{eq-action}.
\end{enumerate}
We refer to \cite{gonzalez2010uo} for the complete proofs of these statements.

\section{Energy-stepping Monte Carlo}
\label{sec-esmc}

With reference to Section \ref{sec-hmc}, the energy-stepping Monte Carlo (ESMC) method can now be set forth as a {\sl Hamiltonian Monte Carlo method that uses the energy-stepping integrator} for generating the proposal distribution. Equivalently, ESMC is a Markov chain Monte Carlo method that proposes new states by integrating \emph{exactly} the approximate Hamiltonian
\begin{equation}
    H_h(q,p) = V_h(q) + \frac{1}{2}p^T M^{-1}p\,,
\end{equation}
corresponding to the Legendre transform of the approximate Lagrangian~\eqref{eq:ES:Lh}. In the same way that the energy-stepping integrator defines a Hamiltonian propagator $\Phi_h$ at every step, ESMC defines a Hamiltonian propagator $\Psi_T : \Omega \times \mathbb{R}^n \to \Omega \times \mathbb{R}^n$ that maps the $k-$th sample of the Markov chain and a random momentum to the $(k+1)-$ sample and a momentum that is discarded. This propagator is the composition map
\begin{equation}
    \Psi_T = \Phi_h\circ\Phi_h\circ\cdots\circ\Phi_h\,,
\end{equation}
with as many updates $\Phi_h$ as steps in the integration and, therefore, it is a indeed Hamiltonian propagator. Given a state $q_k$, the energy-stepping Monte Carlo method advances $q_k$ by sampling a random momentum $p_k$ and defining $q_{k+1} = \psi_T(q_k)$, with
\begin{equation}
\label{eq-esmc-update}
 \psi_T(q_k) = (\Pi_1 \circ \Psi_T) (q_k,p_k)\,,
\end{equation}
$\Pi_1$ being the projection onto the first phase coordinate, and {\sl with no rejection whatsoever}.

The ESMC method, by construction, shares the advantageous properties of HMC. In addition, a remarkable benefit is accrued from its zero-rejection ratio: for the (small) computational cost, namely, the solution of a scalar nonlinear equation per step, which is negligible in high dimensional sample spaces, the effort expended in the proposal stage is never wasted.

\subsection{Some remarks on the step size}
As explained in Section \ref{sec-es}, the energy-stepping method integrates Hamilton's equations in time with a granularity dictated by the energy step~$h$. This is in contrast with classical time integration schemes, such as leapfrog, that use the time step size to control the resolution of the incremental updates. In explicit methods, moreover, this time step size is also bound by the Courant–Friedrichs–Lewy (CFL) stability condition. In practical terms, the time step size and hence the computational cost of the proposal stage are constrained by the variation of the potential energy. Again, in contrast, the energy-stepping integrator is exact for the terraced potential and thus stable for any choice of energy step.

In HMC methods, the accuracy of the proposal step is controlled by the length $T$ of the dynamic excursion that every state undergoes once the momentum is randomly selected (see Section~\ref{subs-hmc-details}). If this motion is to be stably integrated by the leapfrog scheme, the time step size must be chosen to be smaller than a certain global bound. Alternatively, this step size can be adapted by estimating the curvature of the potential energy at every integration step. By contrast, once the energy step size is selected for  ESMC, every proposal motion is integrated --- without the need for any adaptivity --- in a number of steps that depends on the gradient of the potential energy. Hence, regions with steep energy and, thus, steep probability density, are integrated with higher resolution by default.

Owing to the different granularity control of the leapfrog and the energy-stepping methods, it is not possible to compare directly their relative cost and discretization error. However, for every integration interval of length~$T$ an \emph{effective time step} can be calculated for the energy-stepping integrator and used as a basis for comparison, namely,
\begin{equation}
    \Delta t_{e} = \frac{T}{i}\,,
\end{equation}
where $i$ is the number of rectilinear trajectories in state space resolved by the algorithm in the interval $[0,T]$, see Section~\ref{sec-es}. However, it should be carefully noted that the effective time step is an average measure for purposes of comparison only, since the integration of a time interval might encompass long and short rectilinear trajectories precisely due to the intrinsic adaptivity of the method.

\section{Numerical examples}
\label{sec-examples}

This section investigates the properties of ESMC and compares its performance with other MCMC methods. The goal is to assess --- by means of numerical tests --- the salient properties of the method. Some of these properties are expected to be comparable to other HMC methods but with the added benefit of a 100\% acceptance ratio. In addition, we explore the consequences of the \emph{exact} preservation of all symmetries by the energy-stepping integrator. We select examples that range from one-dimensional functions, for which we have closed-form expression, to high-dimensional cases with and without symmetries.

\subsection{One dimensional sampling}

As a first test, we compare ESMC with standard MCMC methods: the random-walk MCMC --- to be referred to as RWMC --- and HMC. We seek to sample from the bi-modal, one-dimensional probability function
\begin{equation}
  \label{eq-ex1-p}
  p(x) \propto
  \frac{3}{\sqrt{2\pi 3^2}} \exp[-\frac{(x+2)^2}{18}] + \frac{1}{4\sqrt{2\pi}} \exp[-\frac{(x-4)^2}{2}]\,.
\end{equation}
For RWMC we employ as proposal distribution a Gaussian with zero mean and variance equal to one. The HMC method implementation employs a leapfrog symplectic integrator with a fixed time step of length 1 and a total integration time $T=10$ per proposal. The ESMC method also solves for time integration periods of length $T=10$ and uses a terraced height of value $h=0.35$ that results in an average time step size $\Delta t_e =0.85$, similar to the one employed in HMC.

\begin{figure}[t]
  \centering
  \includegraphics[width=0.75\textwidth]{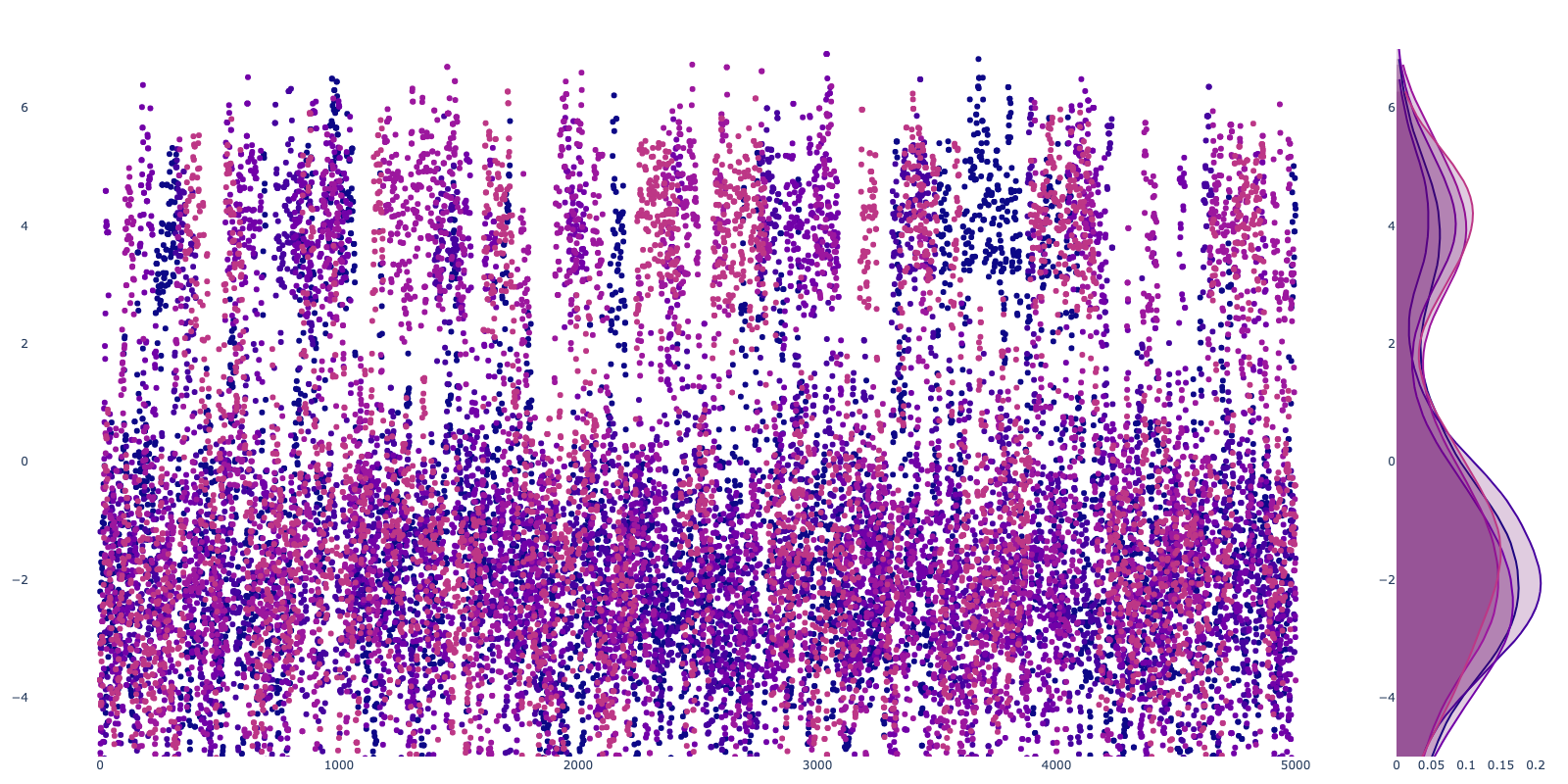}
  \includegraphics[width=0.75\textwidth]{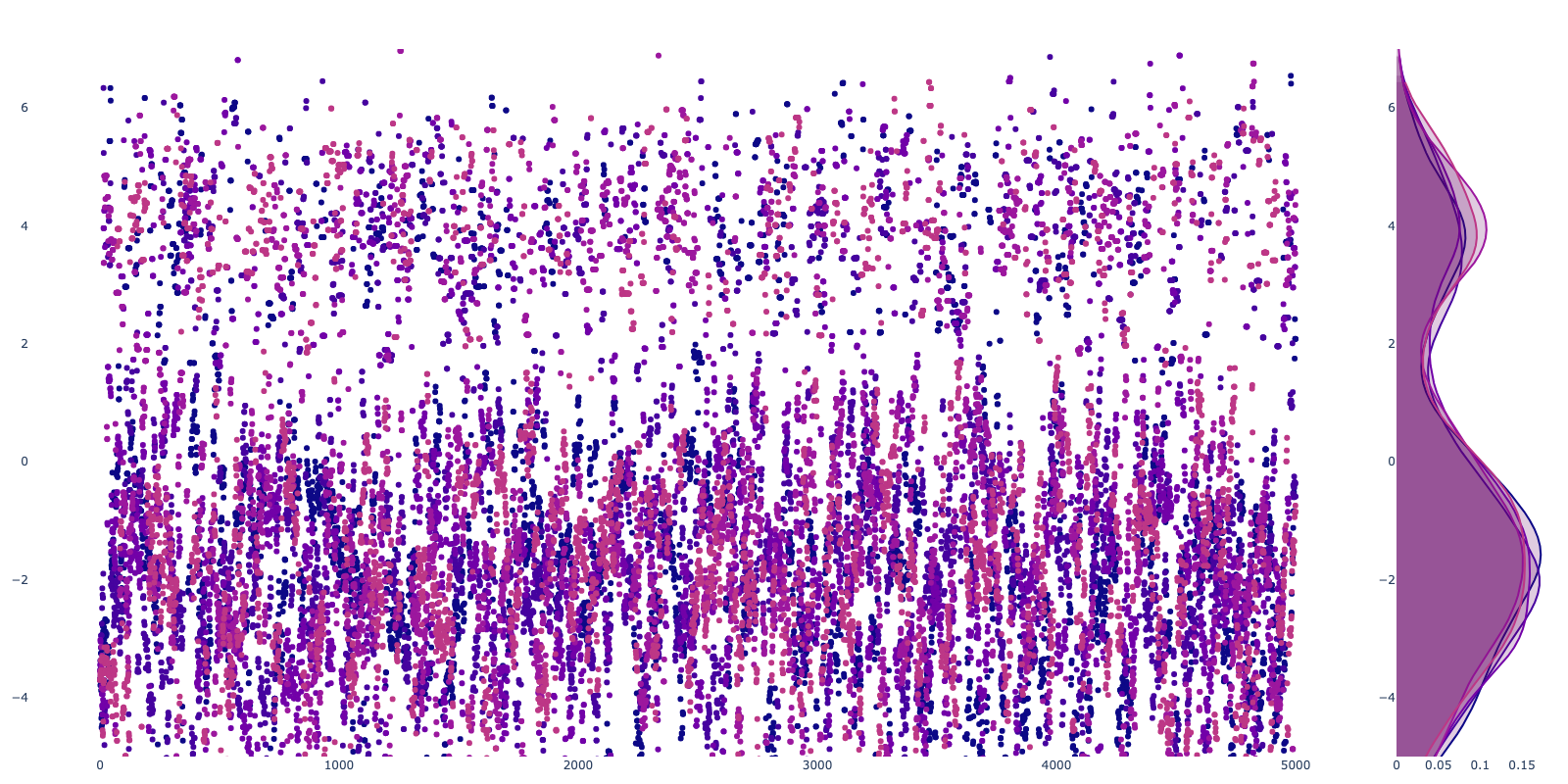}
  \includegraphics[width=0.75\textwidth]{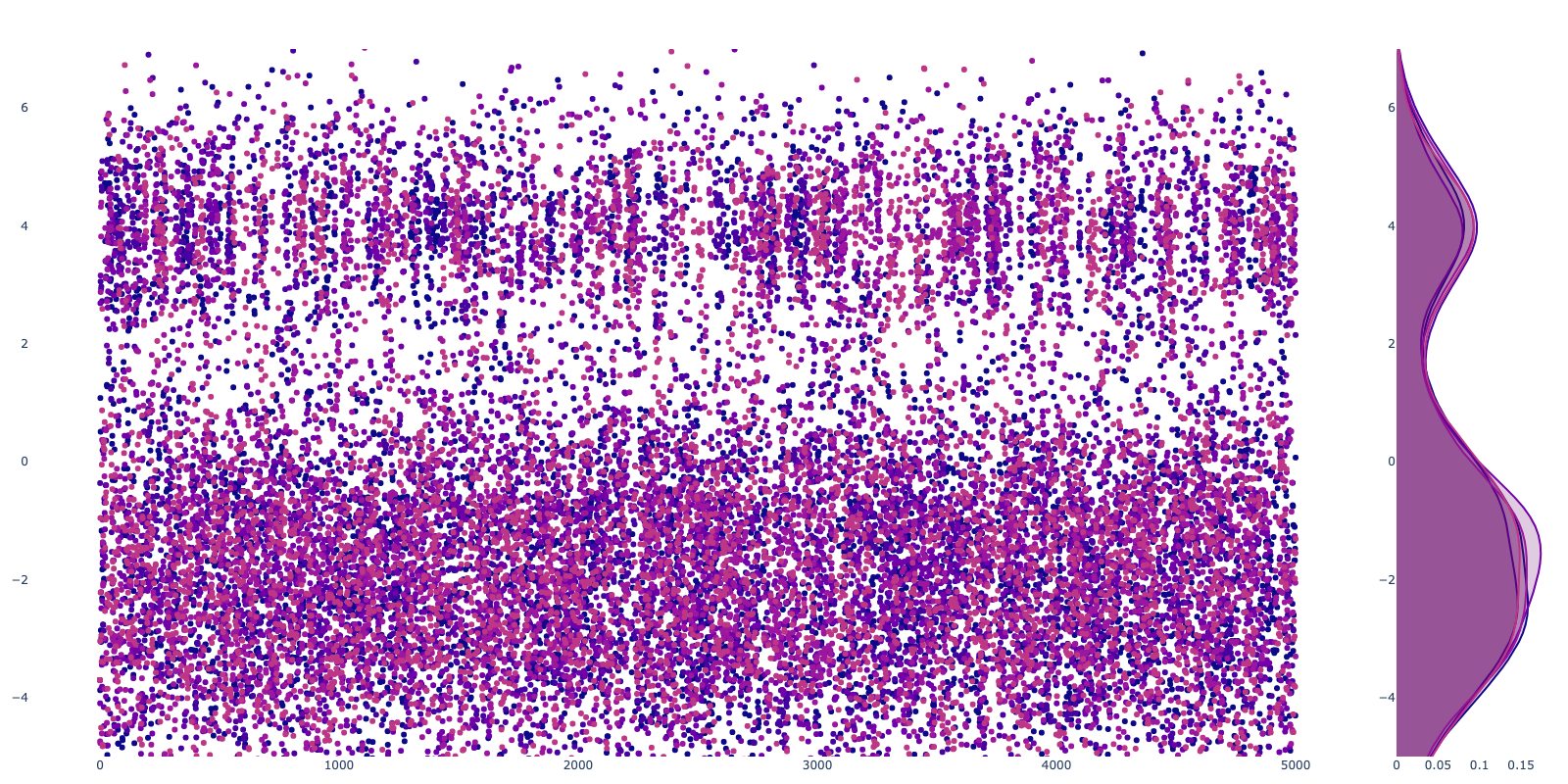}
  \caption{Samples and histograms of for probability~\eqref{eq-ex1-p} obtained with the three MCMC methods. From top to bottom: Random-walk MCMC, Hamiltonian MC, Energy-stepping MC. Five Markov chains are shown for each method.}
  \label{fig-1d}
\end{figure}

To sample the probability $p$, we use Markov chains of length 5000 and a burn-in subset of size 500. For each of the compared methods, we generate five chains and show in Figure~\ref{fig-1d} both the chains as well as the corresponding histograms. In Table~\ref{tab-1d} we summarize the results of the sample generation for the three MCMC methods compared. For each of them, we collect the \emph{average} acceptance ratio across the five chains. Also, for HMC and EMCS we show the total number of integration steps required to produce one chain. Finally, for each of the methods we provide the Kullback-Leibler divergence between the last obtained histogram and the one corresponding to the probability density~\eqref{eq-ex1-p}.

\begin{table}[ht]
  \begin{center}
    \begin{tabular}{ l c  c c  c  }
      & RWMC
      & HMC
      & ESMC\\
      \hline
      Acceptance ratio
      & 0.81
      & 0.72
      & 1.0 \\
      Number of integration steps
      & ---
      & 50000
      & 57530 \\
      KL error
      & 0.04
      & 0.04
      & 0.02
      \\ \hline
    \end{tabular}
  \end{center}
  \caption{Results from sample generation for the bimodal probability distribution~\eqref{eq-ex1-p}. Notation: RWMC: random-walk MCMC, HMC: Hamiltonian MCMC, ESMC: energy-stepping MC.}
  \label{tab-1d}
\end{table}

One of the main advantages of HMC when compared to Metropolis-Hastings MCMC and similar methods is that the former generates Markov chains with less correlated samples. This is beneficial to the exploration of the characteristic set of the probability distribution since weakly correlated samples are allowed to wander the probability space more easily than correlated ones. Samples generated with ESMC, being a modified HMC, are expected to be weakly correlated. The different degrees of correlation among samples in the methods compared could be inferred from Fig.~\ref{fig-1d}.

\begin{figure}[p]
  \centering
  \includegraphics[width=0.70\textwidth]{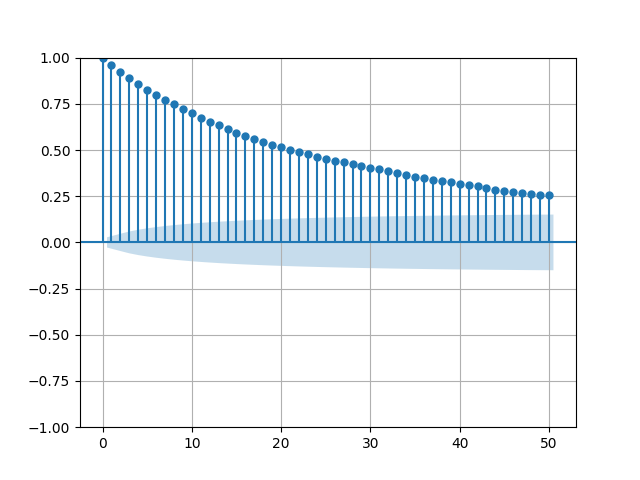}
  \includegraphics[width=0.70\textwidth]{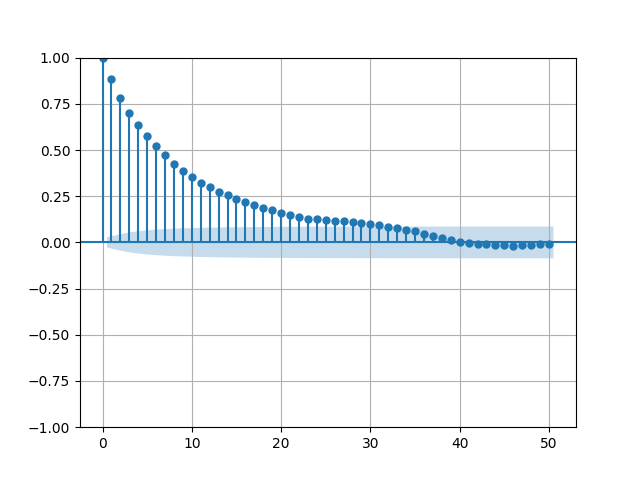}
  \includegraphics[width=0.70\textwidth]{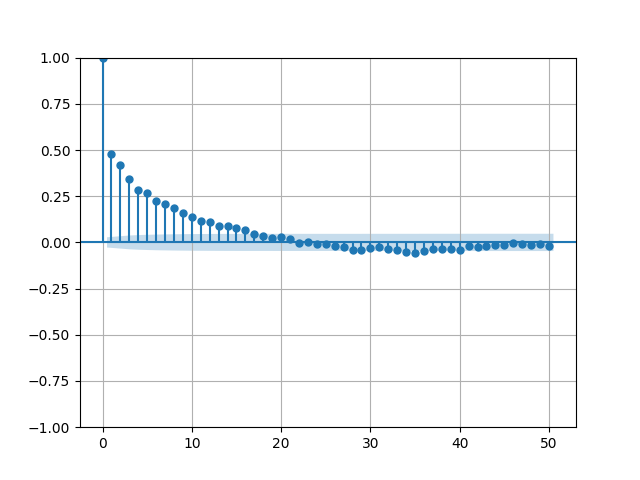}
  \caption{Auto-correlation plots of the Markov chains from example 1 with lags up to 50 samples. From top to bottom: Random-walk MCMC, Hamiltonian MC, Energy-stepping MC.}
  \label{fig-1d-acf}
\end{figure}

A more qualitative measure of sample correlation is provided by the auto-correlation plots of Fig.~\ref{fig-1d-acf}. These figures depict the correlation among samples of one Markov chain, as calculated by each of the three compared methods, all of them with lags smaller or equal to 50. It can be concluded from the figures that the correlation between samples obtained with RWMC is much higher than for samples obtained with HMC and ESMC. One unexpected result, for which we have as yet no clear explanation, is that the correlation of samples in the ESMC chain is also significantly smaller than for the HMC. As Fig.~\ref{fig-1d-acf} shows, the correlation between a sample and the previous one (lag equal to one) is identically one, as a result of the Markov property. However, only in the case of ESMC does this correlation drops abruptly for lags greater than one.

\subsection{Sampling from a mixture Gaussian}
\label{subs-ex2d}
\begin{figure}[ht]
  \centering
  \includegraphics[width=0.7\textwidth]{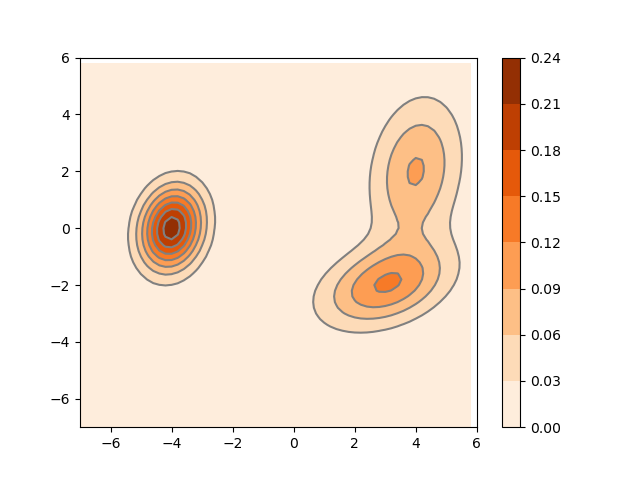}
  \caption{Probability density function for example~\ref{subs-ex2d}.}
  \label{fig-ex2-p}
\end{figure}

As a second example, we further explore the performance of ESMC, now employing it to sample a two-dimensional probability distribution and comparing it with RWMC and HMC. To this end, we now consider the probability distribution $p : \mathbb{R}^2 \to [0,\infty)$
\begin{equation}
  \label{eq-p-2d}
  p(q) \propto
  \sum_{i=1}^3
  \frac{1}{\sqrt{2\pi |\Sigma_i|}} \exp[-\frac{1}{2} (q-\mu_i)\cdot \Sigma_i^{-1}(q-\mu_i)]\,,
\end{equation}
with
\begin{align}
  \label{eq-ex2d-musigma}
    \mu_1 &=
    \begin{Bmatrix}
      4 \\ 2
    \end{Bmatrix}\,,
    & \mu_2 &=
    \begin{Bmatrix}
      3 \\ -2
    \end{Bmatrix}\,,
    & \mu_3 &=
    \begin{Bmatrix}
      -4 \\ 0
    \end{Bmatrix}\,,
    \\
    \Sigma_1 &=
    \begin{bmatrix}
      1 & 1/3 \\ 1/3 & 3
    \end{bmatrix}\,,
    \quad
    & \Sigma_2 &=
    \begin{bmatrix}
      2 & 1/2 \\ 1/2 & 1
    \end{bmatrix}\,,
    \quad
    & \Sigma_3 &=
    \begin{bmatrix}
      1/2 & 1/10 \\ 1/10 & 1
    \end{bmatrix}\,.
\end{align}
The probability $p$ is obtained from the sum of three two-dimensional Gaussian probability distributions with the salient feature that the first two are relatively close, while the measure of the last one is concentrated relatively far from the other two. Thus, the characteristic set of $p$ has two separated regions of high probability (see Fig.~\ref{fig-ex2-p}). Situations such as this one might cause trouble to Metropolis-Hastings-type MCMC methods because the high correlations between samples make it difficult for the Markov chain to leave regions of high probability and explore other characteristic sets. As a result, and depending on the proposal distribution, it might happen that an MCMC method whose initial state falls in or close to a region of high probability would not leave it unless the chain size is large, thus consuming much computational time to sample all relevant regions of sample space.

\begin{figure}[p]
  \centering
  \includegraphics[width=0.5\textwidth]{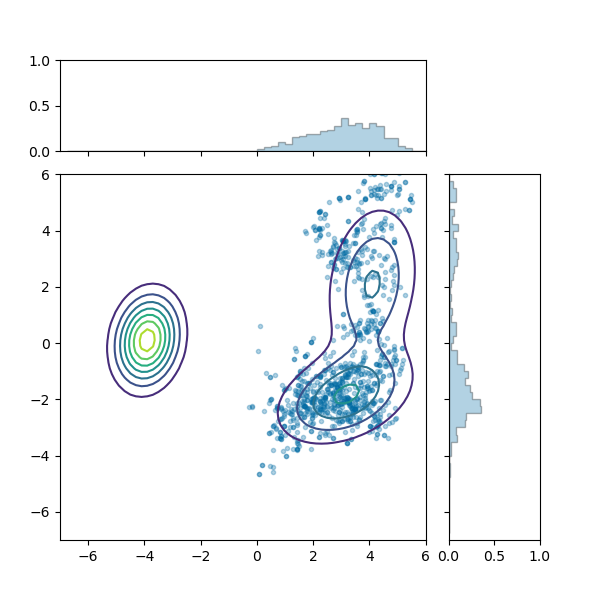}%
  \includegraphics[width=0.5\textwidth]{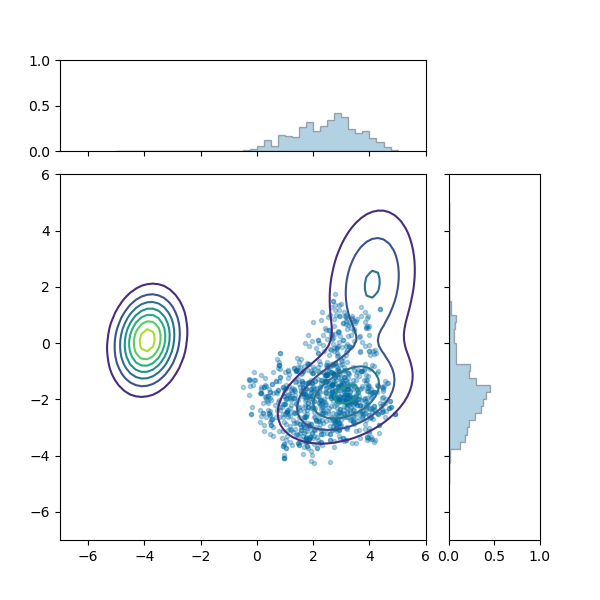}
  \includegraphics[width=0.5\textwidth]{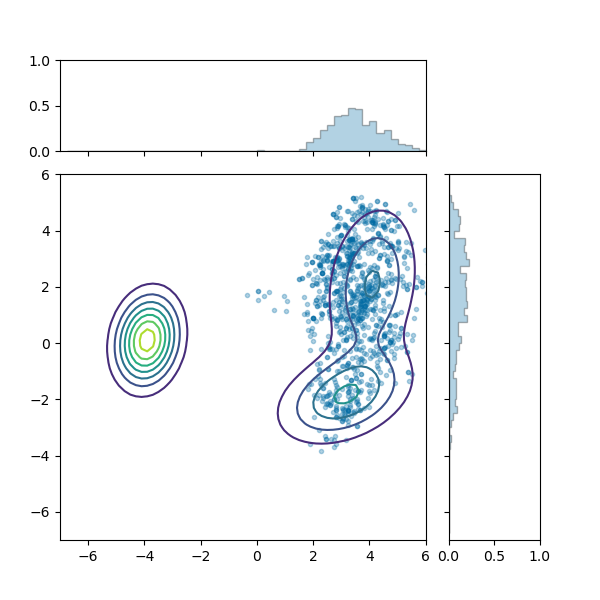}%
  \includegraphics[width=0.5\textwidth]{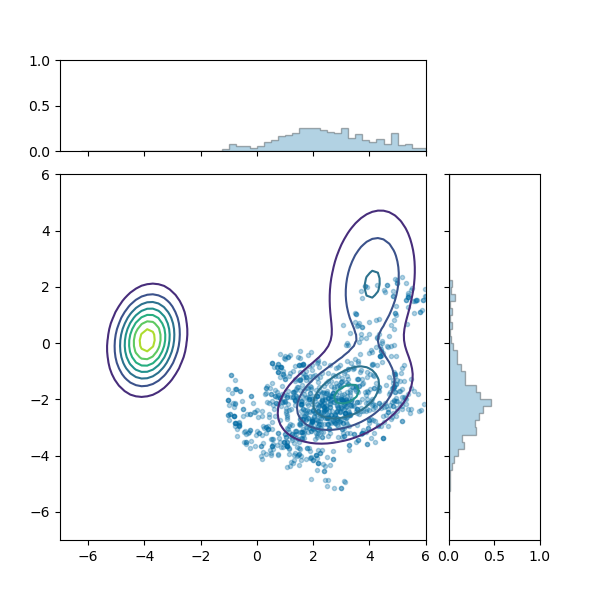}
  \includegraphics[width=0.5\textwidth]{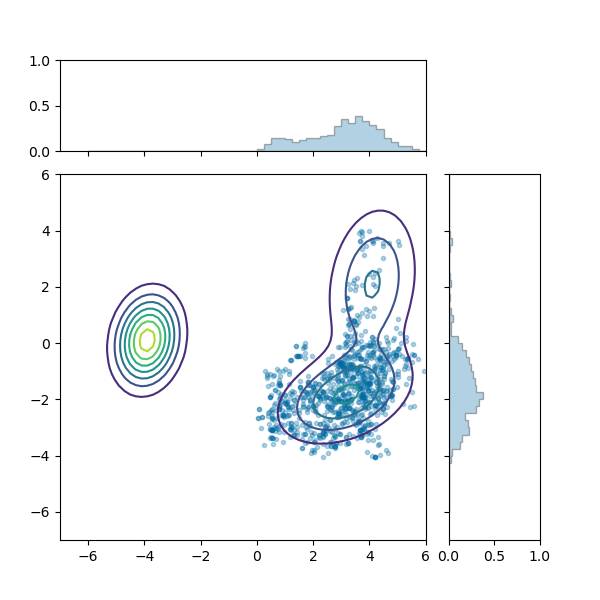}%
  \includegraphics[width=0.5\textwidth]{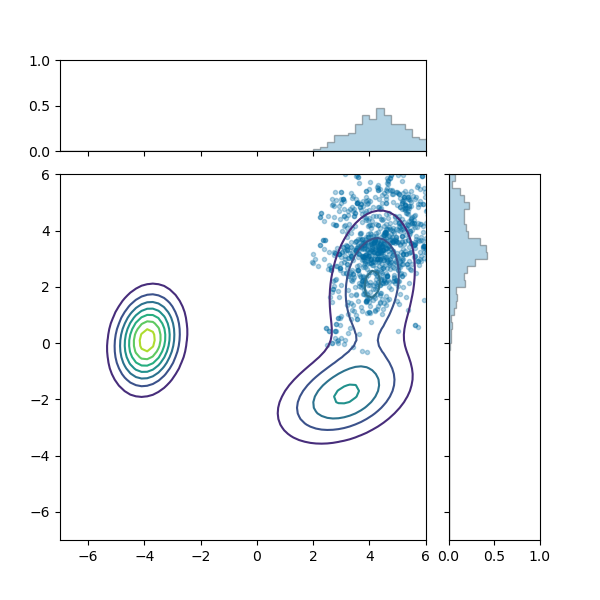}
  \caption{Graphical representation of six (independent) Markov chains for the probability distribution~\eqref{eq-p-2d}, obtained with the random-walk MCMC method. Marginal distributions are plotted at the top and right of each figure.}
  \label{fig-2d-rw}
\end{figure}

\begin{figure}[p]
  \centering
  \includegraphics[width=0.5\textwidth]{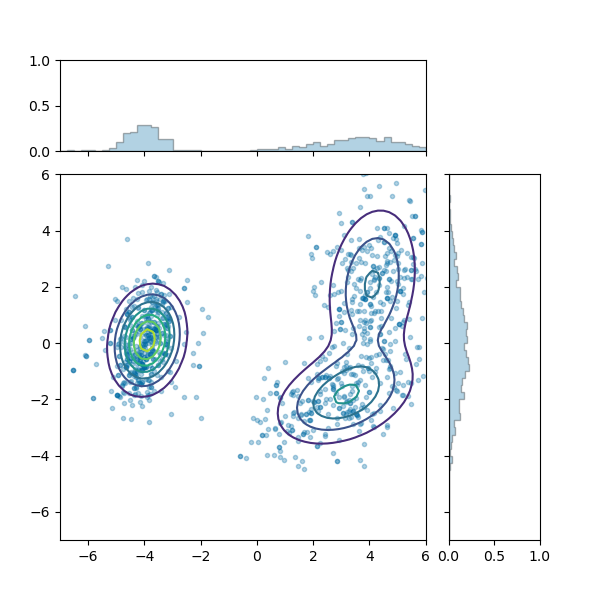}%
  \includegraphics[width=0.5\textwidth]{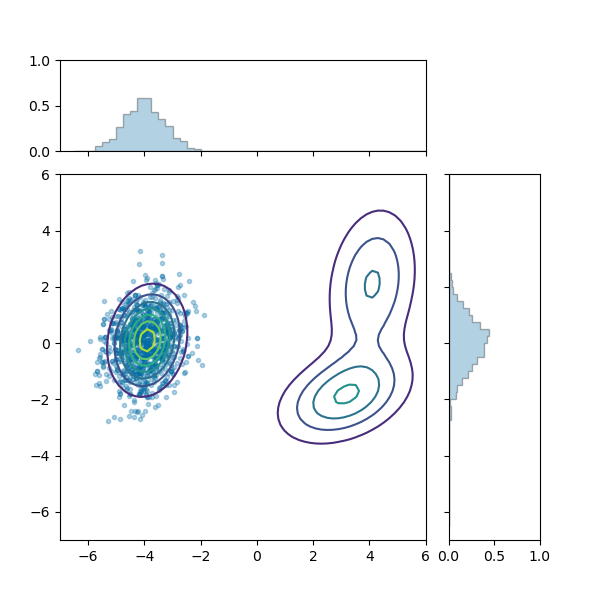}
  \includegraphics[width=0.5\textwidth]{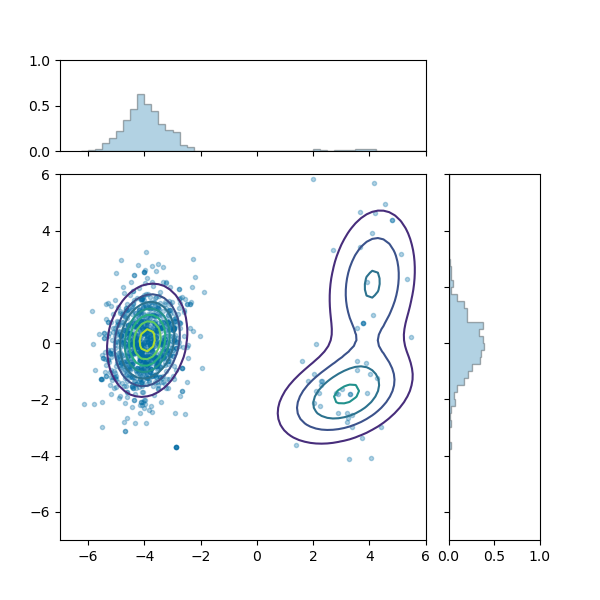}%
  \includegraphics[width=0.5\textwidth]{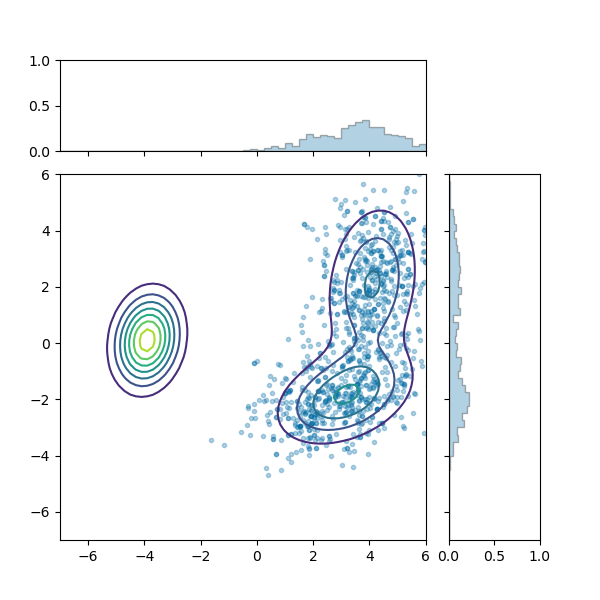}
  \includegraphics[width=0.5\textwidth]{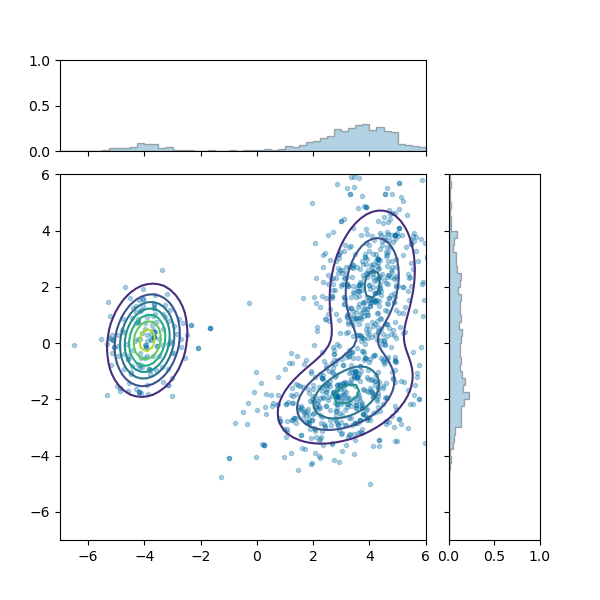}%
  \includegraphics[width=0.5\textwidth]{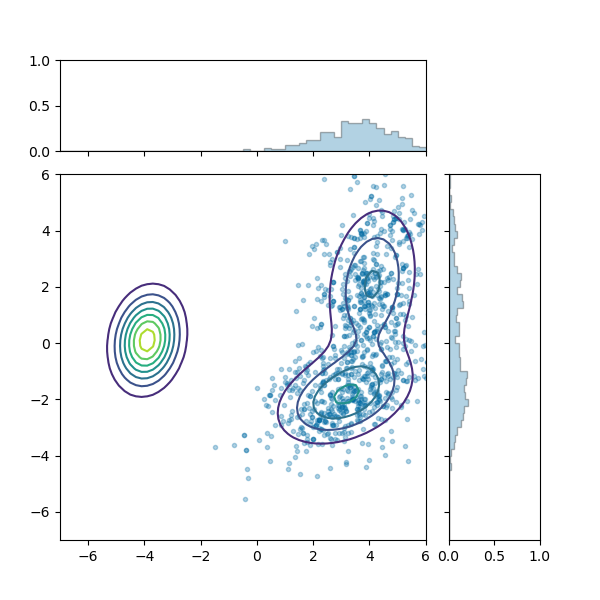}
  \caption{Graphical representation of six (independent) Markov chains for the probability distribution~\eqref{eq-p-2d}, obtained with the Hamiltonian MCMC method. Marginal distributions are plotted at the top and right of each figure.}
  \label{fig-2d-hmc}
\end{figure}

\begin{figure}[p]
  \centering
  \includegraphics[width=0.5\textwidth]{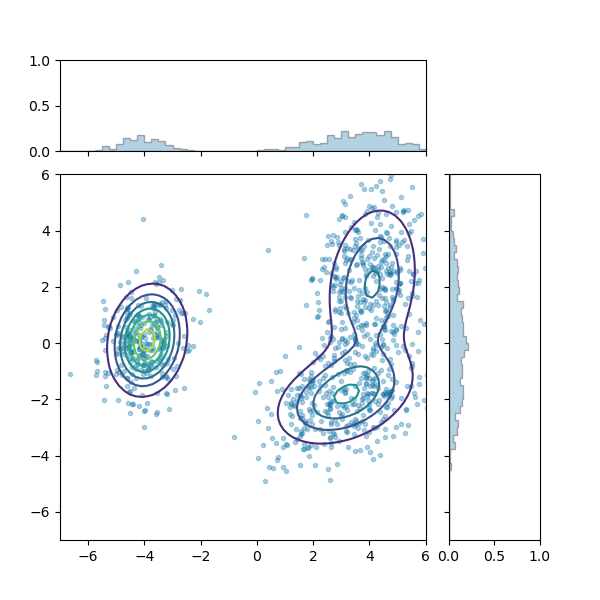}%
  \includegraphics[width=0.5\textwidth]{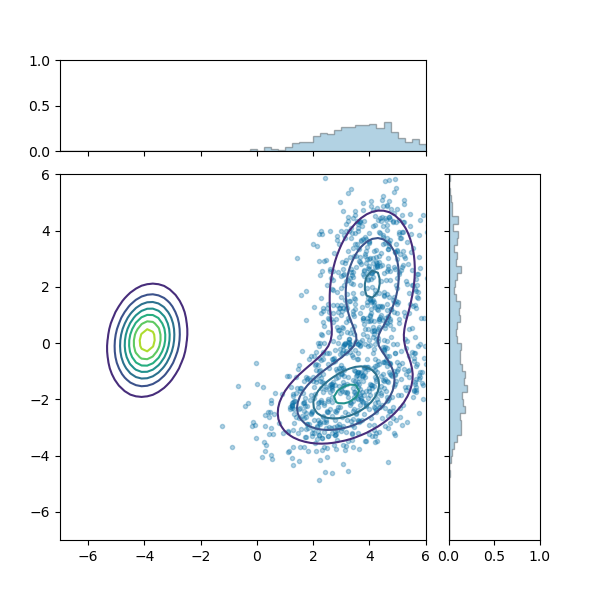}
  \includegraphics[width=0.5\textwidth]{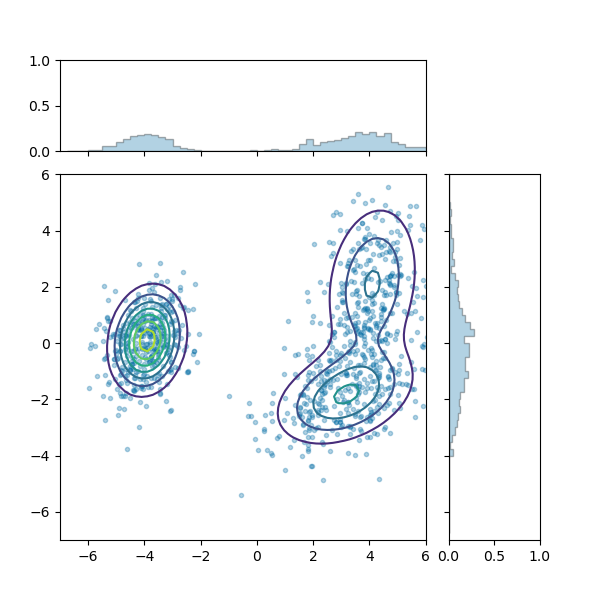}%
  \includegraphics[width=0.5\textwidth]{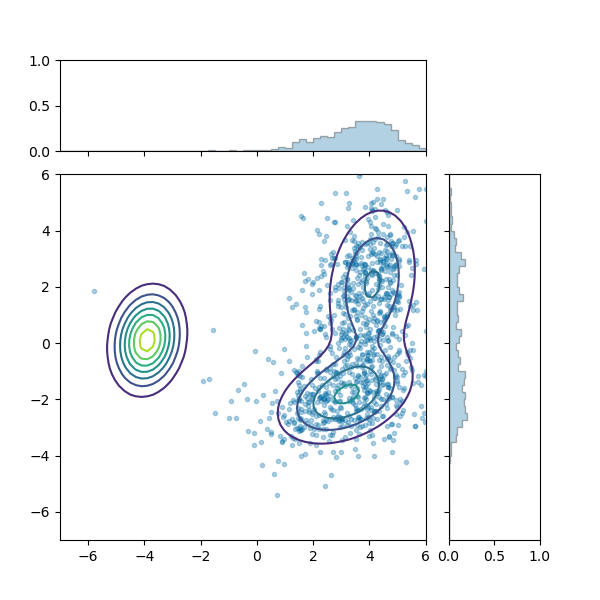}
  \includegraphics[width=0.5\textwidth]{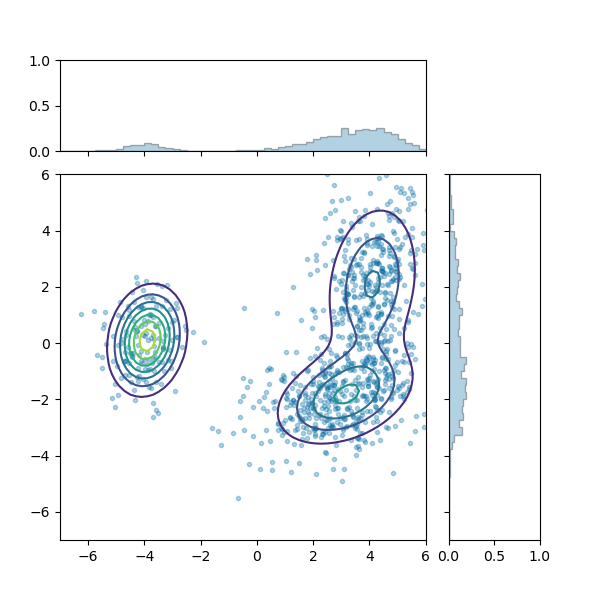}%
  \includegraphics[width=0.5\textwidth]{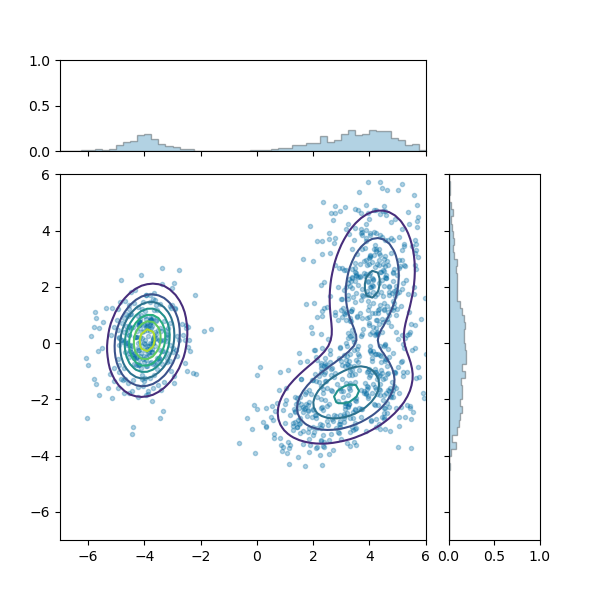}
  \caption{Graphical representation of six (independent) Markov chains for the probability distribution~\eqref{eq-p-2d}, obtained with the energy-stepping MCMC method. Marginal distributions are plotted at the top and right of each figure.}
  \label{fig-2d-esmc}
\end{figure}

To illustrate this effect we simulate next the generation of Markov chains distributed according to~\eqref{eq-p-2d}, using RWMC, HMC, and ESMC. For each of them, we obtain six chains of 1000 samples each and a 10\%\ burn-in ratio, all depicted in Figures~\ref{fig-2d-rw} to~\ref{fig-2d-esmc}. In every solution, the states of the Markov chain are plotted superimposed to the probability density. Additionally, the marginal distributions of every chain are also depicted for easy comparison.

The RWMC solutions are run with a Gaussian proposal probability centered at the previously obtained point and with covariance matrix $\sigma=0.1\, I_{2\times2}$. This variance is chosen to be small relative to the distance between the centers of the three Gaussian functions appearing in Eq.~\eqref{eq-p-2d}; this value is selected to illustrate that, unless specific measures are taken, Metropolis-Hastings type MCMC methods might fail to sample in all the relevant sets of the probability distribution. Figure~\ref{fig-2d-rw} shows six random Markov chains obtained using RWMC, all of them producing samples with an average acceptance ratio of 0.83. Depending on the initial (random) state, the six solutions fail to sample in both sets where the probability measure is concentrated but rather cluster close to the initial state. This behavior depends on the specific choice of proposal distribution employed. If the one employed is replaced with another two-dimensional Gaussian with variance $\Sigma=2\,I_{2\times2}$, the situation is somewhat improved with some chains covering both characteristic sets, at the expense of reducing the acceptance rate to~0.29, on average.

Figures~\ref{fig-2d-hmc} and~\ref{fig-2d-esmc} show, respectively, the samples of six random and independent Markov chains calculated using HMC and ESMC. The simulations obtained with HMC employ time integration periods for the proposal of length 10 and time step size equal to~1. In turn, ESMC obtains its proposals by integrating over periods of the same length, but using instead a terraced potential of step size $h=1$. This step size is selected so that the average time step size employed in generating the proposals is slightly above 1, and almost equal to that selected for HMC. With this choice of parameters, the acceptance ratio for HMC is 0.88.

The results of Figs.~\ref{fig-2d-hmc} and~\ref{fig-2d-esmc} confirm that these two methods produce chains that can explore the two characteristic regions of the probability density~\eqref{eq-p-2d}. In comparison to RWMC, HMC and ESMC can provide more accurate Markov chains for a fixed number of samples. They can exploit the geometric information of the gradient of $p$ to sample more efficiently the space. At the same time, this additional information comes at a higher computational cost.

\subsection{A high-dimensional problem}
Next, we consider an example that has been used in the literature to study the effect of the time integration scheme used in HMC on the acceptance ratio \cite{blanes2014rx,calvo2021hp}. The goal of this test is to sample from a probability function with density
\begin{equation}
\label{eq-blanes-pi}
\pi(q) \propto \exp\left[ - \frac{1}{2} \sum_{k=1}^N k^{2}\,q_k^2 \right] \,,
\end{equation}
the probability density function of a multivariate Gaussian with zero mean and diagonal covariance matrix $\Sigma_{kk}= k^{-2}$. The potential $V(q) = \log(-\pi(q))$ corresponds to the potential energy of $N$ decoupled harmonic oscillators, each of them with stiffness $k^2$.

If the kinetic energy employed in HMC and ESMC is quadratic with a diagonal mass matrix equal to the $N$-dimensional identity matrix, the natural frequencies of the springs would also coincide with $k^2$. Thus, the CFL condition on the leapfrog method restricts the step sizes in the time integration phase of HMC to be $\Delta t<2/N$. We note that ESMC is not subject to such a restriction.

For this problem, we study the acceptance ratio for several dimensions $N$ and also the errors in the sampling. For that, we calculate the sample mean $\bar{\mu}$ and covariance matrix $\bar{\Sigma}$ and define the errors
\begin{equation}
\label{eq-blanes-errors}
    e_1 = \frac{1}{N} \| \bar{\mu} \|\,,
    \qquad
    e_2 = \frac{1}{N}
    \left(
    \sum_{k=1}^N \frac{(\bar{\Sigma}_{kk}-\Sigma_{kk})^2}{\Sigma_{kk}^2}
    \right)^{1/2}
    \,.
\end{equation}
We sample the probability density~\eqref{eq-blanes-pi} using RWMC, HMC, and ESMC in spaces of dimensions $N=2^j$ with $j=2,3,4,5,6$. For the three algorithms, we sample five Markov chains of 5000 states and a 10\%\ burn-in ratio. The initial state of all the samples is obtained from the target distribution, as in \cite{calvo2021hp}.

For the RWMC method, a multivariate Gaussian is selected as the proposal distribution. For each chain, the normal is chosen to be zero and the covariance matrix is the identity matrix scaled by $N^{-2}$. The HMC proposal employs the leapfrog scheme with a time step size $\Delta t=1/N$ in an interval of fixed length~5 (also following \cite{calvo2021hp}). Finally, ESMC is run with an energy step of size $h=\sqrt{N}/2$, which results in effective time step sizes close in value to those in HMC for any given dimension~$N$.

\begin{figure}[t]
  \centering
  \includegraphics[width=0.9\textwidth]{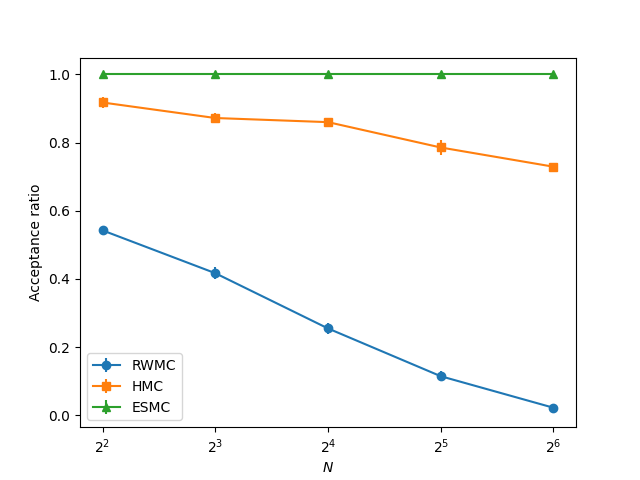}%
  \caption{Acceptance ratio when sampling~\eqref{eq-blanes-pi} using
  the random walk MCMC (RWMC), the Hamiltonian Monte Carlo (HMC) and the energy-stepping Monte Carlo (ESMC).}
  \label{fig-blanes-acceptance}
\end{figure}

\begin{figure}[t]
  \centering
  \includegraphics[width=0.9\textwidth]{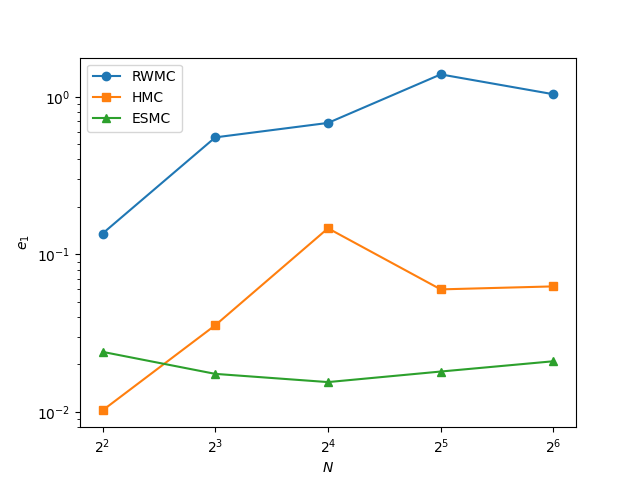}%
  \caption{Error in the sample mean of the Markov chains for probability~\eqref{eq-blanes-pi} using the random walk MCMC (RWMC), the Hamiltonian Monte Carlo (HMC) and the energy-stepping Monte Carlo (ESMC).}
  \label{fig-blanes-mu}
\end{figure}

\begin{figure}[t]
  \centering
  \includegraphics[width=0.9\textwidth]{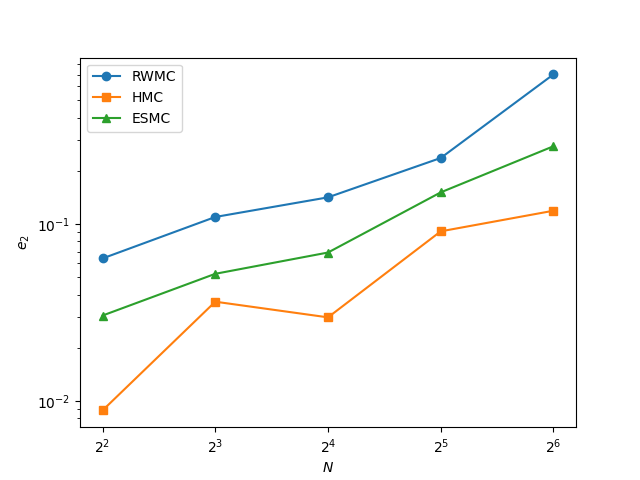}%
  \caption{Error in the sample covariance of the Markov chains for probability~\eqref{eq-blanes-pi} using the random walk MCMC (RWMC), the Hamiltonian Monte Carlo (HMC) and the energy-stepping Monte Carlo (ESMC).}
  \label{fig-blanes-covariance}
\end{figure}

Figure~\ref{fig-blanes-acceptance} shows the acceptance ratio of the three methods compared as a function of the dimension $N$ of the sample space. The acceptance ratio of HMC remains essentially constant for all dimensions. In contrast, the acceptance ratio of RWMC decreases with the dimensionality of the problem. The ESMC method accepts, by construction, all sampled states.

Figures \ref{fig-blanes-mu} and~\ref{fig-blanes-covariance} depict, in logarithmic scale, the values of the errors $e_1$ and $e_2$ defined in Eq.~\eqref{eq-blanes-errors} averaged over the five Markov chains obtained for each method. For all dimensions~$N$, the sampled statistics obtained from HMC and ESMC are closer to the exact values of the target distribution than those obtained with RWMC.

\subsection{Probability distributions with symmetries}
The ESMC method builds upon the energy-stepping integrator and thereby benefits from its favorable properties. For example, the unconditional energy conservation of the integrator is responsible for the zero rejection ratio of the MC proposals. In this next example, we explore the consequences of the symmetry preservation property of the energy-stepping integrator.

As proven in \cite{gonzalez2010hb, gonzalez2010uo}, the energy-stepping integrator preserves all the symmetries of the Lagrangian, even if they are not explicitly identified. This property, together with the exact integration of the approximate Lagrangian, results leads to the exact conservation of momentum maps. However, in the framework of MC methods a symmetry of the potential $V(q)$ alone does not necessarily translate into a symmetry of the Lagrangian and, by extension, into a symmetry of sampling distribution. It bears emphasis that the difficulty arises from the standard assumption of a quadratic kinetic energy with constant mass, which breaks the symmetry of the potential in general. An attractive alternative is to utilize kinetic energies with the same symmetries as the potential, in which case the entire machinery of momentum maps would be recovered by the energy-stepping integrator. This limitations notwithstanding, we may expect the general symmetry conservation properties of the energy-stepping integrator to be qualitatively beneficial.

\begin{figure}[t]
  \centering
  \includegraphics[width=0.49\textwidth]{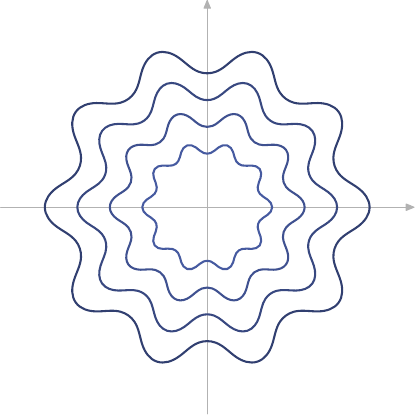}
  \includegraphics[width=0.49\textwidth]{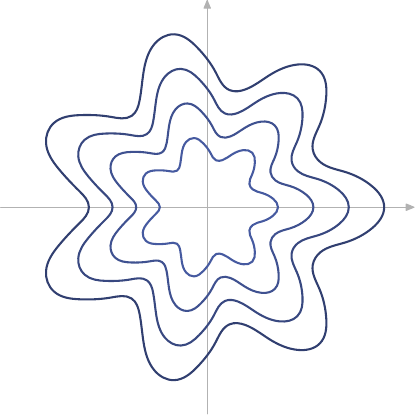}
  \caption{Curves of the family~\eqref{eq-flower}. Left: $\gamma=0.1$~rad, $n=10$; right:
    $\gamma=0.2$~rad, $m=7$.}
  \label{fig-flower}
\end{figure}

To study the consequences of symmetry preservation, we consider an illustrative example involving sampling a probability density function with a non-trivial symmetry. To start, let the angle~$\gamma$ and the integer~$m$ be fixed. Then, for all $\rho\ge0$, consider the closed curves on the plane with polar coordinates $(r,\theta)$ by
\begin{equation}
  \label{eq-flower}
   r(\theta) = \rho\, (1+\sin\gamma \cos(m\,\theta))\,.
\end{equation}
See Figure~\ref{fig-flower} for an illustration. Each pair $(\gamma,m)$ describes a family of flower-shaped curves with $m$ petals, each of them of size proportional to $\sin\gamma$. Within the family, each curve corresponds to one value of the parameter $\rho$ and all possible angles $\theta\in[0,2\pi)$. Hence, if the parameters $(\gamma,m)$ are fixed, every point $\mbs{x}$ of the plane --- except for the origin --- can be mapped onto a unique pair $(\rho,\theta)$ with
\begin{equation}
  \label{eq-flower-coordinates}
  \rho = \hat{\rho}(\mbs{x}) = \frac{r}{1+\sin\gamma\,\cos(m\,\theta)}\,,
\end{equation}
where $(r,\theta)$ are the polar coordinates of $\mbs{x}$. Notice that the coordinate~$\rho$ selects the closed curve in the family where the point lies on, and $\theta$ determines uniquely its position on the curve. Hence, $(\rho,\theta)$ serve as curvilinear coordinates for points on the plane.

\begin{figure}[t]
  \centering
  \includegraphics[width=0.6\textwidth]{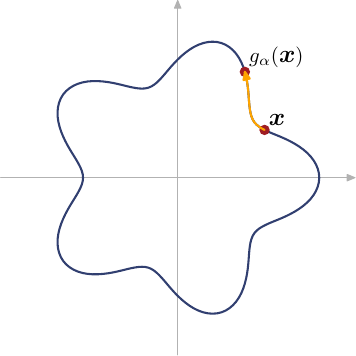}
  \caption{Illustration of the map $g_{\alpha}$ of Eq.~\eqref{eq-flower-symmetry} moving points on a flower-shaped curve.}
  \label{fig-flower-map}
\end{figure}

Given any point $\mbs{x}\in \mathbb{R}^2$, there exists a map $g$ that moves its position on the flower-shaped curve to which it belongs simply by modifying its second curvilinear coordinate
\begin{equation}
  \label{eq-flower-symmetry}
  g: [0,2\pi) \times \mathbb{R}^2  \to \mathbb{R}^2\,,
  \qquad
  g_{\alpha}(\mbs{x}) = \tilde{g}_{\alpha}(\rho,\theta) \mapsto \mbs{y} \equiv (\rho,\theta+\alpha)\,.
\end{equation}
See Figure \ref{fig-flower-map} for an illustration of this map. Maps $g_{\alpha}$, $\alpha\in[0,2\pi]$ can be composed and inverted; in addition $g_0$ is the identity map. Hence, the set $G=\{g_{\alpha},\ \alpha\in[0,2\pi]\}$ is a group of diffeomorphisms that leave all flower-shaped curves invariant.

Next, we turn to the problem of interest, namely the sampling from a probability distribution defined on $\mathbb{R}^2$ and of the form
\begin{equation}
  \label{eq-flower-probability}
   \pi(\mbs{x}) = \tilde{\pi}(\hat{\rho}(\mbs{x}))\,.
\end{equation}
Since the group $G$ leaves the coordinate $\rho$ of all points unchanged, it is a symmetry of the probability $\pi$, that is,
\begin{equation}
  \label{eq-flower-action}
  \pi\circ g = \pi\,,
\end{equation}
for any $g\in G$. It bears emphasis that $G$ is a symmetry of $\pi$ also when the latter is a function of the Cartesian coordinates $(x,y)$, as long as the probability can be written as in Eq.~\eqref{eq-flower-probability}.

We wish to ascertain whether the use of a symmetry-preserving integration scheme like the energy-stepping method has any beneficial effects when sampling a (symmetric) probability distribution such as $\pi$. More specifically, we use Markov chain methods to obtain samples of $\pi$ on the plane, without using any explicit reduction of the probability function or projection onto the quotient space $\mathbb{R}/G$.

Specifically, we consider a probability distribution of the form
\begin{equation}
  \label{eq-flower-ex-p}
  \tilde{\pi}(\rho) \propto \exp[-\frac{1}{2} \rho^2]\,,
\end{equation}
and proceed to sample directly on $\mathbb{R}^2$ using the composed map~\eqref{eq-flower-probability}. For our tests, we select a family of flower-shaped curves with $\gamma=1/3$ and $m=15$. Then, RWMC, HMC and ESMC are compared using samples of size 1000, 2000, 3000, 4000, and 5000 points with a burn-in of 10\%. In all cases, the initial samples of the Markov chains are obtained from a binormal distribution centered at the origin with identity covariance matrix. The RWMC method is run with a proposal distribution that is also a binormal centered at the origin, but with covariance $\Sigma=0.1\,I_{2\times2}$. The HMC method is solved with a time step of 0.3 for intervals of length 0.3. The ESMC method employs a potential terrace of step $h=1$, selected so that the effective time step size is close to~0.3.

\begin{figure}[p]
  \centering
  \includegraphics[width=0.49\textwidth]{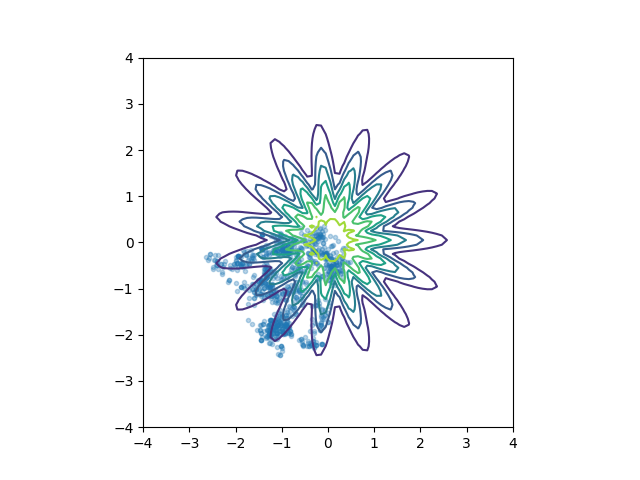}
  \includegraphics[width=0.49\textwidth]{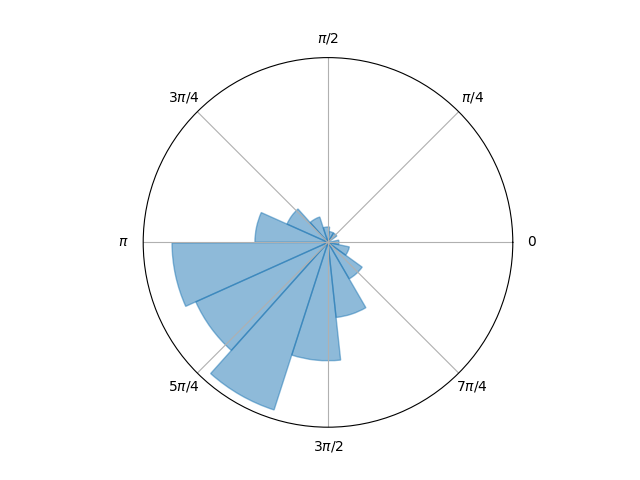}
  \includegraphics[width=0.49\textwidth]{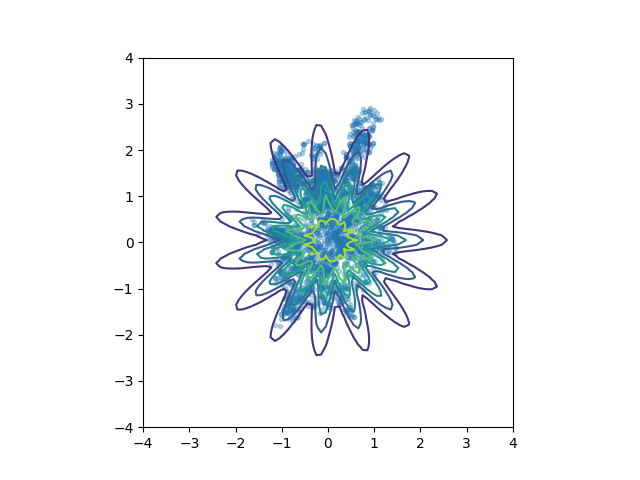}
  \includegraphics[width=0.49\textwidth]{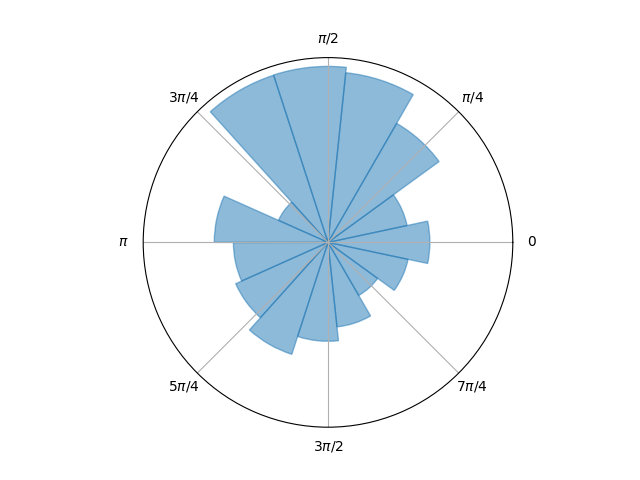}
  \includegraphics[width=0.49\textwidth]{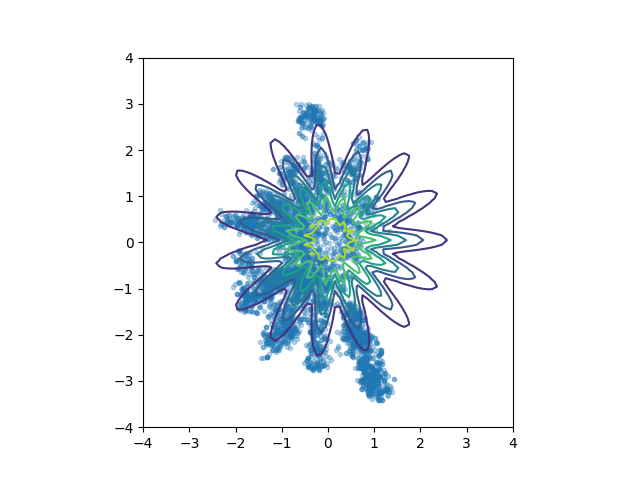}
  \includegraphics[width=0.49\textwidth]{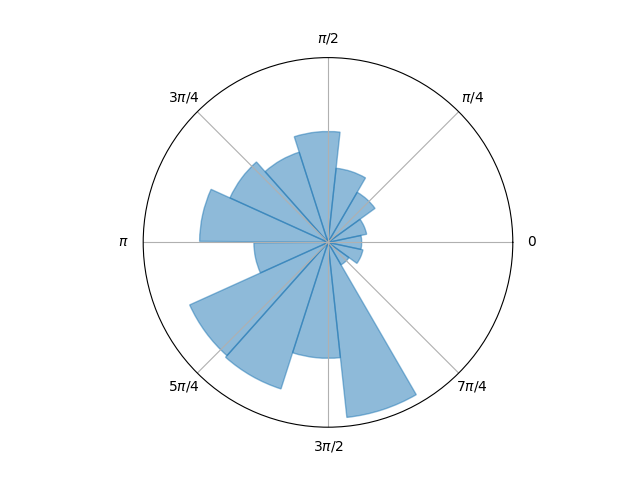}
  \caption{Sampling probability density function~\eqref{eq-flower-ex-p} ($\gamma=1/3, m=15)$ with RWMC. From top to bottom: chains of 1000, 3000, 5000 samples. Left column: scatter plot; right column: histogram of samples binned according to the angular sector.}
  \label{fig-inv-rw}
\end{figure}

\begin{figure}[p]
  \centering
  \includegraphics[width=0.49\textwidth]{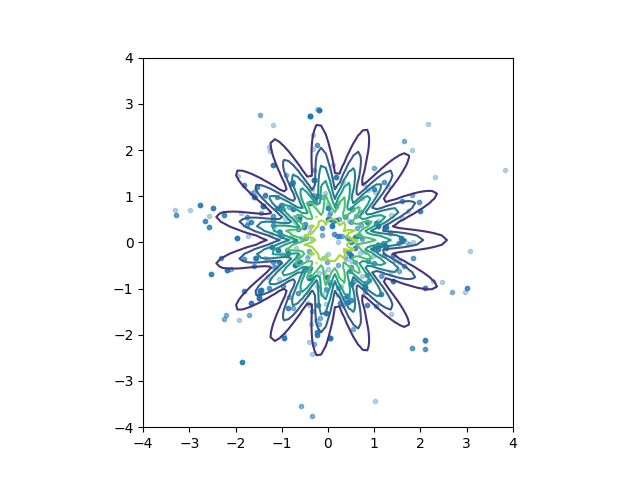}
  \includegraphics[width=0.49\textwidth]{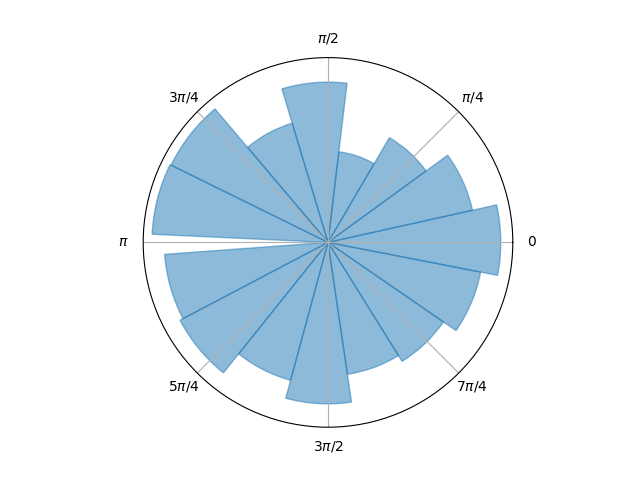}
  \includegraphics[width=0.49\textwidth]{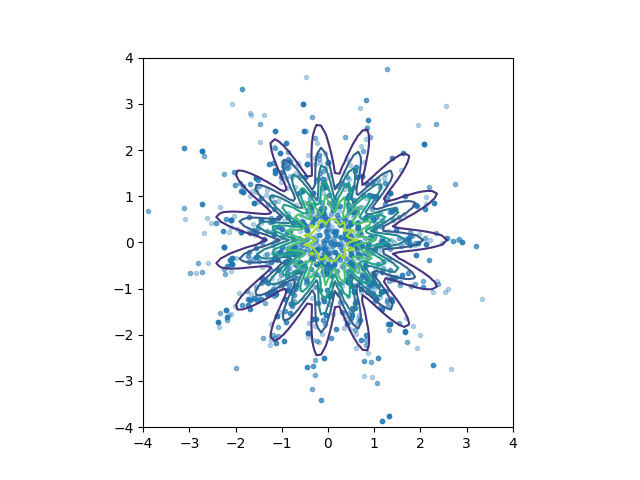}
  \includegraphics[width=0.49\textwidth]{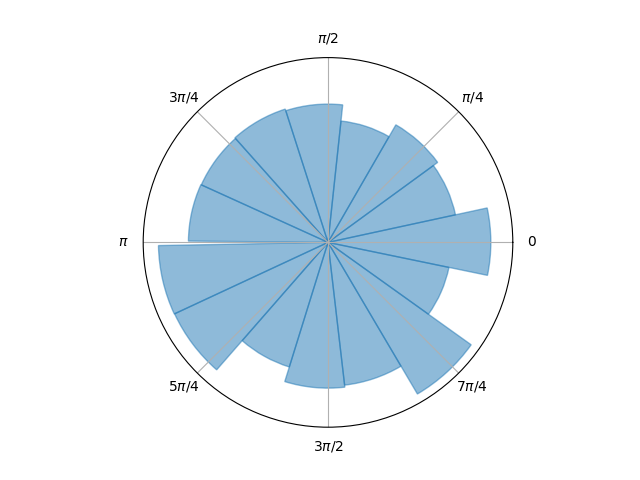}
  \includegraphics[width=0.49\textwidth]{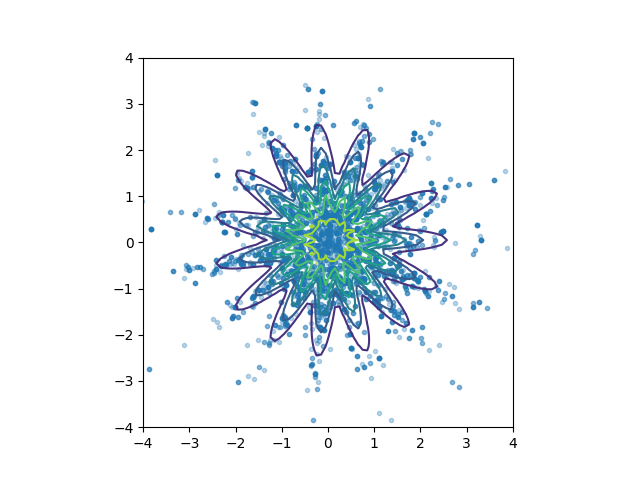}
  \includegraphics[width=0.49\textwidth]{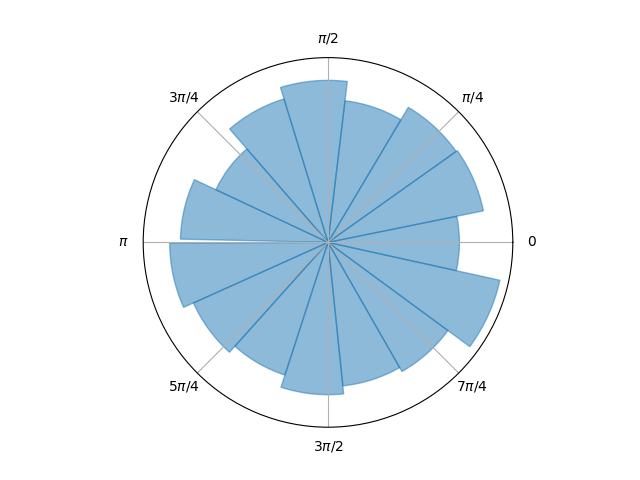}
  \caption{Sampling probability density function~\eqref{eq-flower-ex-p} ($\gamma=1/3, m=15)$ with HMC. From top to bottom: chains of 1000, 3000, 5000 samples. Left column: scatter plot; right column: histogram of samples binned according to angular sector.}
  \label{fig-inv-hmc}
\end{figure}

\begin{figure}[p]
  \centering
  \includegraphics[width=0.49\textwidth]{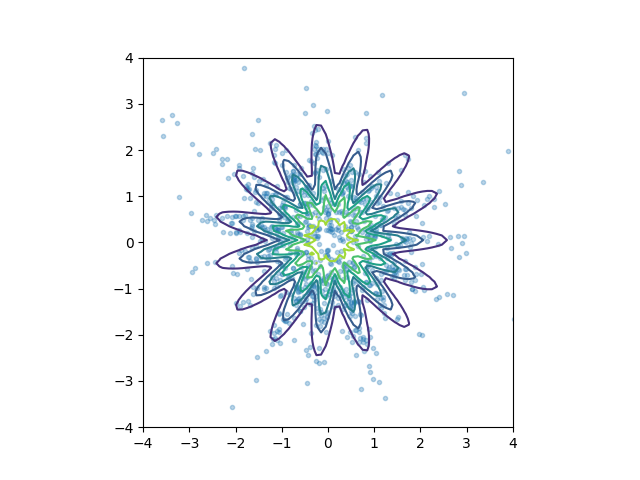}
  \includegraphics[width=0.49\textwidth]{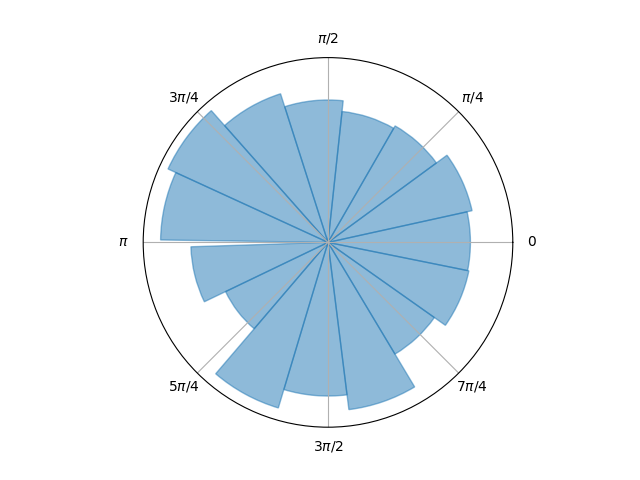}
  \includegraphics[width=0.49\textwidth]{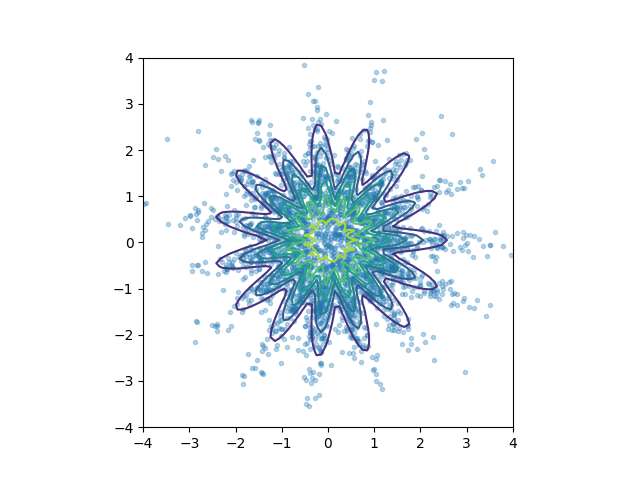}
  \includegraphics[width=0.49\textwidth]{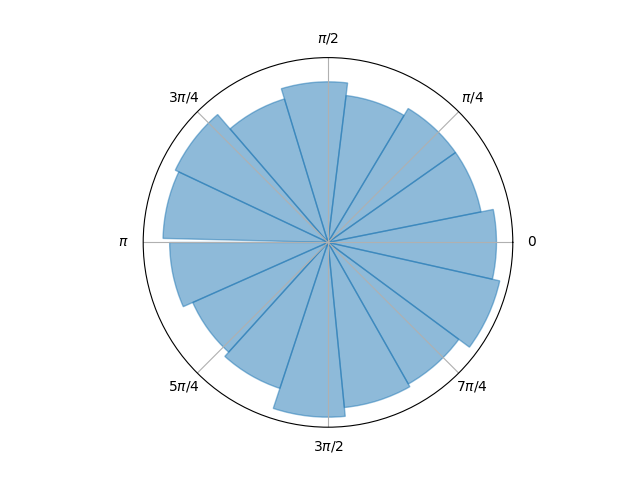}
  \includegraphics[width=0.49\textwidth]{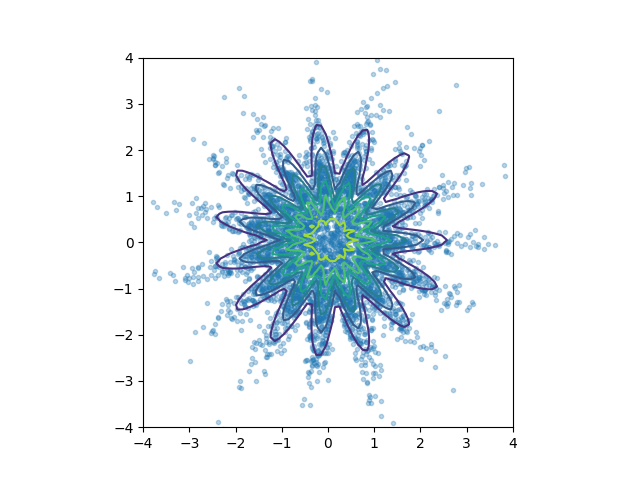}
  \includegraphics[width=0.49\textwidth]{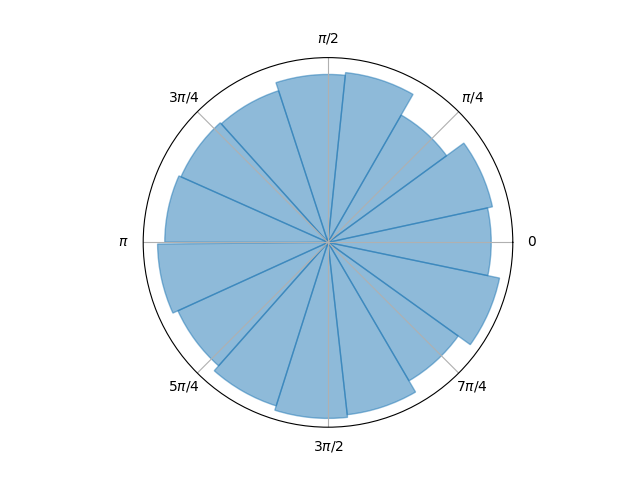}
  \caption{Sampling probability density function~\eqref{eq-flower-ex-p} ($\gamma=1/3, m=15)$ with ESMC. From top to bottom: chains of 1000, 3000, 5000 samples. Left column: scatter plot; right column: histogram of samples binned according to angular sector.}
  \label{fig-inv-esmc}
\end{figure}

The results of the Monte Carlo samples for probability~\eqref{eq-flower-ex-p} are shown in Figures~\ref{fig-inv-rw}, \ref{fig-inv-hmc}, and \ref{fig-inv-esmc} for RWMC, HMC, and ESMC, respectively. On the left of this figure, scatter plots of the accepted samples (without the burn-in) in the chains are depicted. On the right of each scatter plot, a histogram is shown with all the samples distributed among $m=15$ bins, each of them collecting the samples on identical circular sectors centered at the origin. A sampling method that \emph{exactly} preserved the symmetry~\eqref{eq-flower-symmetry} would give histograms where all the bins contained the same amount of samples. Due to the Markovian nature of all MCMC methods, the strict verification of such property is not to be expected.

The first apparent fact from the three figures~\ref{fig-inv-rw}, \ref{fig-inv-hmc}, and \ref{fig-inv-esmc} is that the number of accepted samples in HMC is much larger than in RWMC. Moreover, ESMC accepts \emph{all} samples of the chain, so the number of samples in the scatter plots and histogram is even larger than in the HMC case, for all chain lengths.

Figure~\ref{fig-inv-rw} shows the results obtained with RWMC. The scatter plots show a strong lack of symmetry of the sampled chains. We observe that once a chain falls in the probability basin of one petal, it tends to stay in the corresponding angular sector. This effect is apparent in the histograms of the right column of Figure~\ref{fig-inv-rw}. By increasing the covariance of the proposal distribution employed for this method, this lack of isotropy could be reduced; this strategy, however, would have deleterious effects on the acceptance ratio of the method.

The results of HMC are presented in Figure~\ref{fig-inv-hmc}. Both from the scatter plots and the histograms it can be concluded that the method preserves better than RWMC the symmetry of the probability function. Finally, the chains obtained from ESMC are plotted in Figure~\ref{fig-inv-esmc}. Not only the number of acceptance data points in each chain is significantly larger than for the other two methods compared, but also the histograms reveal that the symmetry is also better preserved.

\begin{figure}[t]
  \centering
  \includegraphics[width=0.8\textwidth]{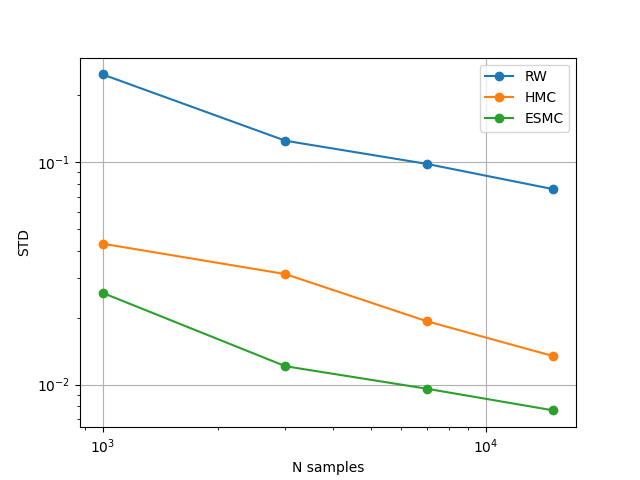} \caption{Standard deviation of the binned data for the RWMC, HMC, and ESMC methods for chains of lengths between 1000 and 15000 samples.}
  \label{fig-inv-std}
\end{figure}

Figure~\ref{fig-inv-std} collects information about the symmetry of the sampled chains. A sampling method that \emph{exactly} preserved the symmetry of the probability would produce samples that are identically distributed per unit angle. Figure~\ref{fig-inv-std} depicts, as a function of the chain length, the standard deviation of the polar angle of all the samples in each chain. Since the probability~\eqref{eq-flower-probability} is independent of the polar angle (i.e., it is symmetric with respect to the action of $G$) the samples should be distributed identically across all angles and the standard deviation of the probability per unit angle should be zero. Figure~\ref{fig-inv-std} shows that HMC and ESMC have a much smaller standard deviation than RWMC and, as expected from the histograms of Figures~\ref{fig-inv-hmc} and \ref{fig-inv-esmc}, the variance of the ESMC chains is always smaller than that of the HMC chains.

The numerical results shown in this section are obtained from one single chain of each method, for each sample size. However, the conclusions are representative of the general behavior of the compared methods. In all simulation runs, the RWMC method fails to preserve the symmetry of the probability distribution, especially for short chains. The HMC and ESMC methods perform much better in this respect, but ESMC always produces chains that are more uniformly distributed in the angular coordinate. As mentioned at the beginning of this section, we speculate that this advantageous feature is a consequence of the \emph{exact} symmetry preservation of the energy-stepping integrator, although proof of this statement is lacking at present.

\section{Summary and conclusions}
\label{sec-summary}
Markov chain Monte Carlo (MCMC) methods consist of the iterative evaluation of two steps: first, a new sample is proposed; then it is stochastically determined if this sample is to be accepted or rejected. The Hamiltonian Monte Carlo method (HMC) exploits a dynamical interpretation of the proposal step to efficiently explore the sample space, covering the characteristic set of the sampled distribution more efficiently than, for example, random-walk MCMC. A vital ingredient of HMC is the numerical integration of Hamilton's equations to generate a proposal state. This is commonly done employing the leapfrog method, an explicit, symplectic algorithm with remarkable qualitative properties. The energy error in the time integration of the leapfrog is used in the acceptance/rejection step of HMC. Here, a stochastic rule determines that proposals with large energy errors are more likely to be rejected.

In this work, we have introduced the energy-stepping Monte Carlo (ESMC) method, an HMC method that replaces the leapfrog scheme with the energy-stepping integrator, a symplectic, quasi-explicit, exact energy-conserving time integration method that, when used in the context of mechanical systems, preserves all the symmetries of the Lagrangian. Owing to the energy conservation property of the energy-stepping integrator, ESMC does not reject any proposals and explores the sample space more efficiently than other existing methods, irrespective of the granularity of the numerical approximation. The numerical tests presented provide empirical evidence that ESMC affords some additional benefits: the Markov chains it generates have weak autocorrelation and it yields smaller errors than chains sampled with HMC and similar time step sizes. Finally, ESMC benefits from the symmetry conservation properties of the energy-stepping integrator when sampling from potentials with built in symmetries, whether explicitly known or not.

In closing, we emphasize that the numerical examples presented in this paper are strictly academic and not intended for a direct head-on performance comparison with highly-tuned libraries such as STAN and NIMBLE. Indeed, owing to their inordinate importance, MCMC methods and HMC, in particular, have been extensively developed algorithmically over the course of decades. As a result, several notable improvements have been proposed (e.~g.~\cite{hoffman2011ns}) resulting in exceedingly efficient implementations. However, is should be possible to further develop ESMC algorithmically in order to bring its performance more in line with the current production codes. This and other enhancements of the method suggest themselves as worthwhile avenues for further research.

\section*{Acknowledgements}
I.~R. has been partially supported by funding received from project OPTIMATED from the Spanish Ministry of Science and Innovation (Proj. no. PID2021-128812OB-I00). M.~O. gratefully acknowledges the support of the Deutsche Forschungsgemeinschaft (DFG, German Research Foundation) {\sl via} project 211504053 - SFB 1060; project 441211072 - SPP 2256; and project 390685813 -  GZ 2047/1 - HCM.

\section*{Supplementary material}
A \texttt{Python} implementation of RWMC, HMC, and ESMC can be downloaded from the public repository \texttt{git@gitlab.com:ignacio.romero/esmc.git}. In addition to the Markov chain methods, the link includes scripts to run all the examples of Section~\ref{sec-examples}.

\bibliography{biblio}
\bibliographystyle{unsrt}

\end{document}